\documentclass[acmsmall]{acmart}

\AtBeginDocument{%
  }

\setcopyright{acmcopyright}
\copyrightyear{2023}
\acmYear{2023}
\acmDOI{XXXXXXX.XXXXXXX}

\acmJournal{TOIS}
\acmVolume{37}
\acmNumber{4}
\acmArticle{111}
\acmMonth{8}

\usepackage{xspace}
\usepackage{algorithmic}
\usepackage{algorithm2e}
\usepackage{multirow}

\usepackage{amssymb}
\usepackage{caption} 
\usepackage[normalem]{ulem}
\usepackage{subfigure}
\usepackage{makecell}
\newcommand{\eat}[1]{}

\newcommand{\baby}{\textsc{LLM-Rec}\xspace}
\newcommand{\Rmnum}[1]{\uppercase\expandafter{\romannumeral #1}}

\begin{document}
\title{One Model for All: Large Language Models are Domain-Agnostic Recommendation Systems}
\author{Zuoli Tang$^*$}
\email{tangzuoli@whu.edu.cn}
\affiliation{%
  \institution{Key Laboratory of Aerospace Information Security and Trusted Computing, School of Cyber Science and Engineering, Wuhan University}
  \country{China}
}

\author{Zhaoxin Huan}
\email{zhaoxin.hzx@antgroup.com}
\affiliation{%
  \institution{Ant Group}
  \country{China}
}

\author{Zihao Li}
\email{zihao.li@whu.edu.cn}
\affiliation{%
  \institution{Key Laboratory of Aerospace Information Security and Trusted Computing, School of Cyber Science and Engineering, Wuhan University}
  \country{China}
}

\author{Xiaolu Zhang}
\email{yueyin.zxl@antfin.com}
\affiliation{%
  \institution{Ant Group}
  \country{China}
}

\author{Jun Hu}
\email{zhaoda.hj@antgroup.com	}
\affiliation{%
  \institution{Ant Group}
  \country{China}
}

\author{Chilin Fu}
\email{chilin.fcl@antgroup.com	}
\affiliation{%
  \institution{Ant Group}
  \country{China}
}

\author{Jun Zhou}
\email{jun.zhoujun@antfin.com}
\affiliation{%
  \institution{Ant Group}
  \country{China}
}

\author{Lixin Zou$^{\dagger}$}
\email{zoulixin@whu.edu.cn}
\affiliation{%
  \institution{Key Laboratory of Aerospace Information Security and Trusted Computing, School of Cyber Science and Engineering, Wuhan University}
  \country{China}
}

\author{Chenliang Li$^{\dagger}$}
\email{cllee@whu.edu.cn}
\affiliation{%
  \institution{Key Laboratory of Aerospace Information Security and Trusted Computing, School of Cyber Science and Engineering, Wuhan University}
  \country{China}
}

\thanks{
$^{\dagger}$ Corresponding author. \\
$^*$ Work done when Zuoli Tang was an intern at Ant Group. \\
$^{\ddagger}$ The source code is available at https://github.com/WHUIR/LLMRec
}

\renewcommand{\shortauthors}{Z. Tang et al.}

\begin{abstract}
Sequential recommendation systems aim to predict users’ next likely interaction based on their history. However, these systems face data sparsity and cold-start problems. Utilizing data from other domains, known as multi-domain methods, is useful for alleviating these problems. However, traditional multi-domain methods rely on meaningless ID-based item representation, which makes it difficult to align items with similar meanings from different domains, yielding sup-optimal knowledge transfer. This paper introduces \baby, a framework that utilizes pre-trained large language models (LLMs) for domain-agnostic recommendation. Specifically, we mix user's behaviors from multiple domains and concatenate item titles into a sentence, then use LLMs for generating user and item representations. By mixing behaviors across different domains, we can exploit the knowledge encoded in LLMs to bridge the semantic across over multi-domain behaviors, thus obtaining semantically rich representations and improving performance in all domains. Furthermore, we explore the underlying reasons why LLMs are effective and investigate whether LLMs can understand the semantic correlations as the recommendation model, and if advanced techniques like scaling laws in NLP also work in recommendations. We conduct extensive experiments with LLMs ranging from 40M to 6.7B to answer the above questions and to verify the effectiveness of \baby in multi-domain recommendation$^{\ddagger}$.
\end{abstract}



\received{25 December 2023}
\received[revised]{25 May 2024}
\received[revised]{29 September 2024}
\received[accepted]{7 November 2024}

\maketitle

\section{Introduction}
\label{sec:intro}
\begin{figure}[!]
    \centerline{\includegraphics[width=0.8\textwidth]{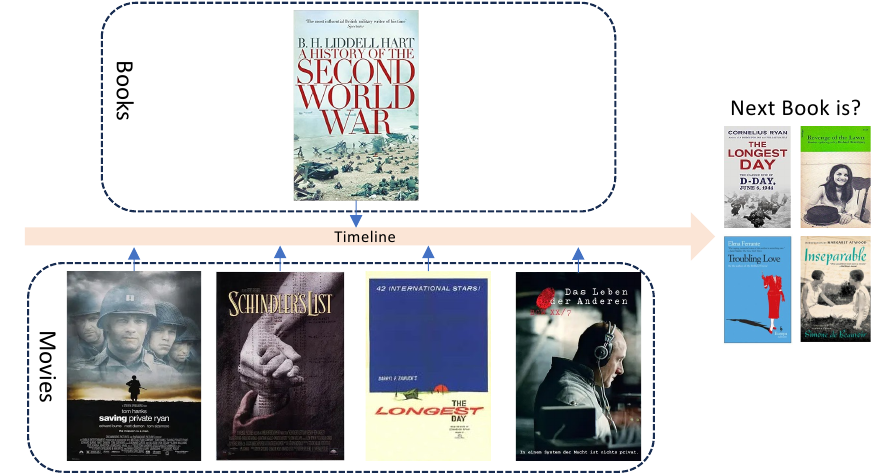}}
    \caption{Users’ interests across various domains exhibit semantic correlations. A user who enjoys war-themed movies may also be interested in books related to World War II.}
    \label{fig:intro_1}
\end{figure}
Based on the historical interactions, sequential recommendation aims to capture users' interests and provides accurate suggestions among massive candidate items to address the information overload problem. It has been widely used in various e-commerce and advertising companies.
Although the previous sequential recommendation methods have achieved great success with the prosperity of deep learning, they still face two challenges: data sparsity and cold-start, which hinder the further development of the recommendation system.
Considering that users' interests across various domains may exhibit semantic correlations, as shown in Figure ~\ref{fig:intro_1}, many research works attempt to introduce user interactions from other domains as auxiliary knowledge to enhance the model's capability (i.e., cross-domain recommendation) and achieve promising results.
Some works ~\cite{PPGN, BiTGCF, H3Trans} construct a unified graph for different domains, and utilize graph neural network to propagate cross-domain information in the graph. Additionally, other works ~\cite{ADI, Star} consider each domain as a task, multi-task approaches(such as MMoE~\cite{MMoE}, PLE~\cite{PLE}) are utilized to solve this problem.

\begin{figure}[!]
    \centerline{\includegraphics[width=0.5\textwidth]{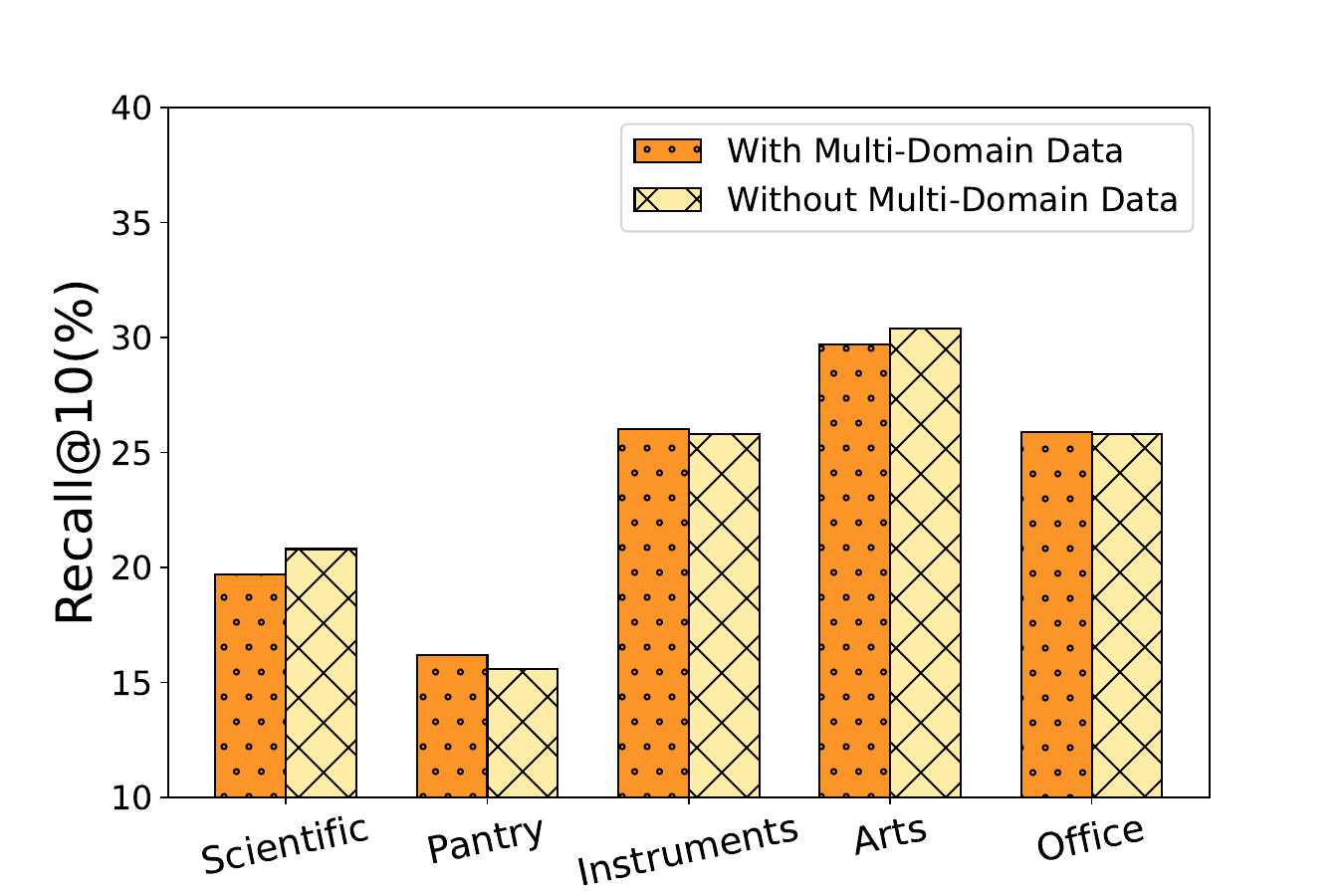}}
    \caption{Applying single-domain ID-based model (SASRec) to multi-domain scenario.}
    \label{fig:intro_2}
\end{figure}

Although applying cross-domain methods becomes ubiquitous in sequential recommendation~\cite{SCDR_1, SCDR_2, MIFN, DDGHM}, there are few works that extend them for multi-domain sequential recommendation since the following obstacles remain to be overcome. 
First, to realize the knowledge transfer between two domains, a complicated model structure is devised in most of the cross-domain solutions. However, the sophisticated pair-wise designs diminish the feasibility and the flexibility of model extension and adaption. 
For example, C$^2$DSR~\cite{SCDR_2} establishes links between two domains through a contrastive objective. Expanding this to multi-domain contexts would introduce additional $A_N^2$ objectives (N means the domain number), complicating the training process significantly. MIFN~\cite{MIFN} designs a unit for measuring the degrees of connection to transfer knowledge between two domains. While the degrees become difficult to quantify due to the $C_N^2$ possible inter-domain relationships in the multi-domain scenario.
Second, the premise assumption of cross-domain recommendation (i.e., the source domain is always better-informed than the target domain, thus, cross-domain recommendation aims to exploit the source domain for target domain improvement) specializes in target domain enhancement while neglecting the source domain. On the contrary, we believe both the source domain and target domain are important.
Last but not least, most of the existing works leverage the ID, e.g., a number code, for item symbolization. We argue that such a string of numbers could not release any transferable information corresponding to items and domains. 
In the preliminary study, we concatenate user interactions across multi-domains straightforwardly in terms of item id. These interaction sequences are then fed into a standard sequential baseline SASRec. Then, we verify the model performance on each single domain. The results are shown in Figure~\ref{fig:intro_2}. We could find that simply merging multi-domain behavior does not bring performance improvement and may even lead to performance degradation. 

Recently, large language models have delivered outstanding capacity in learning world knowledge from text data. We believe the world knowledge encapsulated in LLM could bridge the gap between different scenarios. To this end, we endeavor to leverage the contextual information of each item for multi-domain recommendation. A general domain-agnostic framework with pre-trained LLMs is proposed (namely \baby). More concretely, for each item, we use the item title for item representation generation. As for each individual, we collect a user's interacted items from different domains along the timeline. Then, we further concate the titles of these items as a sentence, and feed them into the LLM for user behavior modeling. By configuring different kinds of LLMs for \baby, we aim to investigate the following questions:

\textbf{Q1:} Is it feasible to use a language model as a unified user encoder and item encoder to model user behaviors across different domains? Can it leverage the advantages of language models to enhance performance on all domains and does cross-domain behaviors contribute to performance improvement? To answer this question, we collect the same user's behaviors across five domains for modeling and observe the resultant performance.

\textbf{Q2:} How does the large language model work for text-based recommendation? More specifically, when employing large language models to encode user interaction sequence information and item text information, does it mainly depend on embedded semantic knowledge to establish the connection between a user's interests as we expect, or does it also model item-level collaborative filtering information? To answer this question, we classify items based on their popularity and analyze the performance of items under different popularity levels. We also analyze the exposure rate of the resultant recommendations. Specifically, when the large language models can well capture the semantic correlation underlying the user's sequential behaviors, the performance and exposure of cold-start items will be greatly reduced. Furthermore, we conduct a visualization study of the item representations to investigate this question further.

\textbf{Q3:} Are the latest techniques related to language models in NLP
still applicable to the recommendation domain?  "Bigger is better" is a common principle in pre-trained models (i.e., in general, the performance will be improved with larger model size in CV~\cite{Vit-22b} and NLP~\cite{FLAN-T5}), but similar conclusions have rarely been studied in recommendations.
To answer these questions, we investigate the performance of three types of language models covering both encoder-only and decoder-only structures. The parameter size of those models ranges from $40$ million to $6.7$ billion. Furthermore, we also exploit the performance of different tuning algorithms. That is, we conduct experiments with LoRA fine-tuning~\cite{Lora} methods on various model sizes and compare its performance to the full-parameter counterpart.

In a nutshell, we make the following contributions in this paper: 
\begin{itemize}
    \item We explore the ability of language models in multi-domain behaviors modeling and demonstrate the effectiveness of world knowledge embedded in language models for multi-domain recommendation.

    \item We conduct comprehensive experiments on five real-world datasets to verify the effectiveness of our method over eight baselines for multi-domain sequential recommendation. Moreover, we conduct a thorough analysis of the role of language models in sequential recommendation.

    \item We further investigate the impact of the critical factors (including \textit{model architectures}, \textit{model sizes}, \textit{data settings}, as well as \textit{fine-tuning methods}) of the large language model to the final performance for the multi-domain sequential recommendation.
\end{itemize}

\section{Related Work}
\label{sec:relwork}
This section provides a brief overview of representative efforts relevant to our work. 
\eat{
in the fields of sequential recommendation, multi-level interest learning, and pre-training recommendation models, and comprehensively compares them with our proposed method.
}

\subsection{Sequential Recommendation}
In the early stage, the dominant efforts were devoted to modeling users' sequential behaviors as a Markov chain such that the item-item transition matrix is learnt for next item prediction~\cite{FMC}. Attributed to the extraordinary capability of deep learning for complicated pattern modeling, a surge of deep neural networks, e.g., RNN-based~\cite{GRU4Rec} and CNN-based~\cite{Caser}, are validated successively for sequential recommendation.
However, these solutions often fail to capture long-term dependency between any of two items in sequence, thus, Transformer-based solutions, e.g., SASRec~\cite{SASRec}, BERT4Rec~\cite{BERT4Rec}, are applied and achieved promising results in sequential recommendation.
Moreover, some sophisticated methods, e.g., contrastive learning~\cite{CL4SRec, ICLRec}  are also adopted to relieve the data sparaity in sequential recommendation.

\subsection{Multi-domain Recommendation}
Although conventional methods achieve remarkable success for many recommendation tasks, there are two long-standing obstacles that hinder the further improvement of recommendation system: data sparsity and cold-start problems. To alleviate this issue, some works~\cite{ADI,MAMDR,BiTGCF} propose to exploit the interaction behaviors from other domains as external information for better performance, namely multi-domain recommendation.
For instance, MCF~\cite{MCF} extracts collaborative information from different domains for recommendation. PPGN~\cite{PPGN} models the user's multi-domain interaction records as a graph. Then, the multi-domain information are aggregated by the graph neural network.
BiTGCF~\cite{BiTGCF} devices a domain feature propagation layer with GNN for knowledge transferring across different domains.
ADI~\cite{ADI} denotes multi-domain recommendation as multi-task learning, then proposes a domain-share network for multi-domain recommendation.
On the contrary, MAMDR~\cite{MAMDR} argues using a domain-share network straightforwardly for multi-domain recommendation will hinder the model to achieve optimal results due to the domain conflicts. Hence, a Domain Negotiation strategy with regularization is proposed to balance the specificity of each domain as well as the commonality among different domains.
However, all the above works require complex model structural design and do not specialize in sequential recommendation, neither of them explores language models for multi-domain recommendation.

\subsection{Text-based Recommendation}
Encouraged by the remarkable success of LLM, very recently, some methods endeavor to apply language models for recommendation~\cite{UniSRec, Recformer, Miracle, MoRec, ZESRec}.
MoRec~\cite{MoRec} explores the feasibility of using textual 
representations (instead of ID embeddings) generated by language model for recommendation.
ZESRec~\cite{ZESRec} explores the zero-shot ability of sequential recommendation model. To be specific, they first select one domain and adopt a pre-trained language model to generate item representations in terms of their associated textual information. The resultant representations are then used for sequential model pre-training. Afterward, the well-trained model will be applied for other domain recommendation.
Following ZESRec, UniSRec~\cite{UniSRec} utilizes multi-domain data to pretrain the sequential model first. Then the downstream domain data is further used for model fine-tuning to obtain better performance.
Very recently, Miracle~\cite{Miracle} combines multi-interest learning with pre-training and proposes a novel sparse capsule routing network.  
Different from~\cite{ZESRec, UniSRec}, Recformer~\cite{Recformer} pre-trains and fine-tunes a language model in a holistic approach for item text encoding and sequential recommendation.
However, these works ~\cite{ZESRec, UniSRec, Recformer} follow the pre-training and fine-tuning, a two-stage paradigm. We believe this recipe might be inefficient and cumbersome to some extent. 
Moreover, none of them investigate the potentiality of language models by taking the items from different domains to form a single user behavior sequence, nor do they explore the effect of model size to the final recommendation performance.

\section{Methodology}
\begin{figure*}
    \centering
    \includegraphics[width=0.95\textwidth]{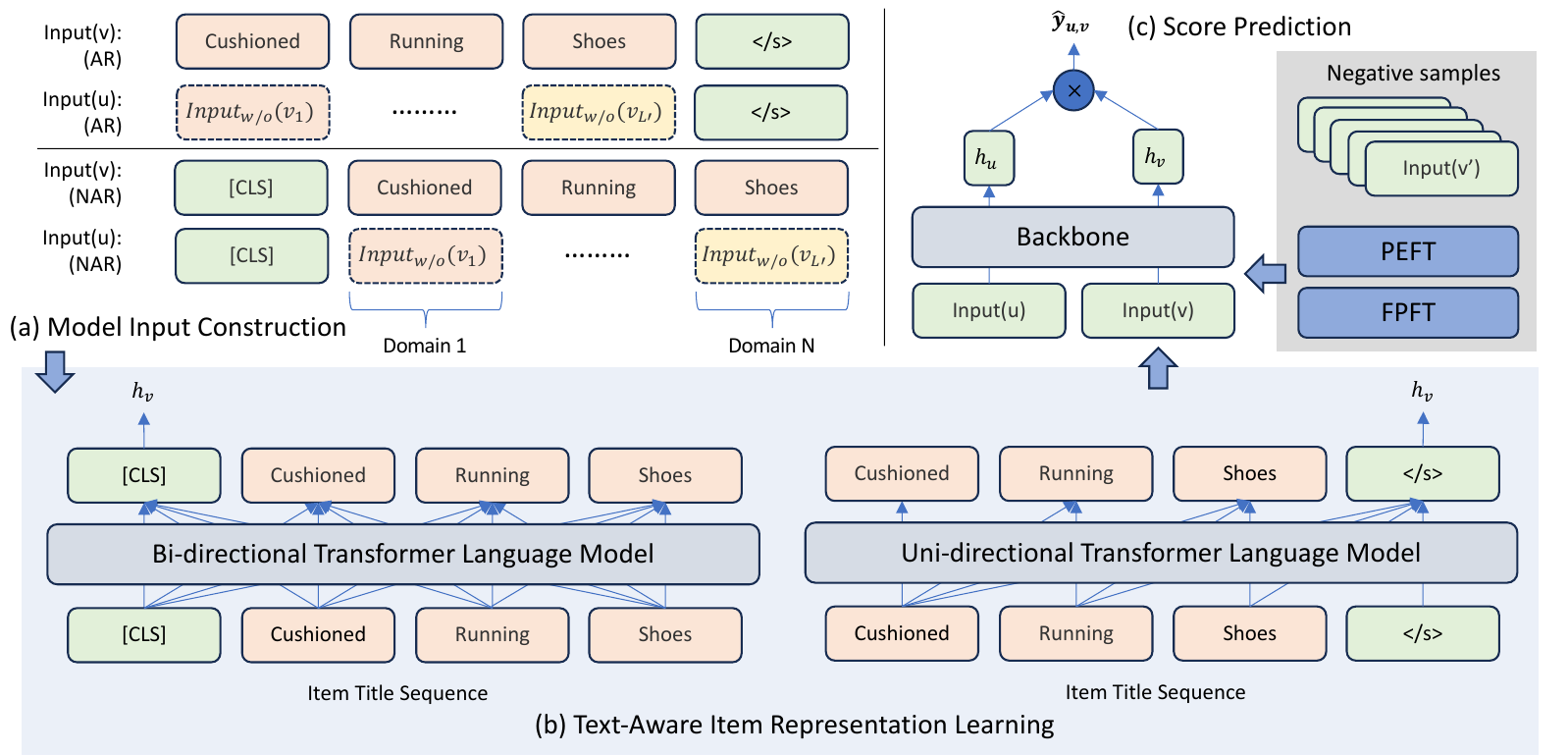}    
    \caption{The overview of the proposed \baby}
    \label{fig:framework}
\end{figure*}

This section outlines the multi-domain sequential recommendation problem and introduces the key components of the proposed \baby: different backbones, model input construction, score prediction, and model optimization. The overall framework is shown in Figure~\ref{fig:framework}.
\label{sec:preliminary}

\subsection{Problem Formulation}
\label{subsection:Problem Formulation}
Given $N$ domains (i.e., $\mathcal{D}^1$, $\mathcal{D}^2$,..., $\mathcal{D}^N$), denoting $\mathcal{U}^i$, $\mathcal{V}^i$ and $\mathcal{R}^i$ as user set, item set and interaction set of domain $i$, respectively. Thus, we could collect the user, item and interactions from all the domains, and let $\mathcal{U}=\mathcal{U}^1\cup \mathcal{U}^2\cup .... \mathcal{U}^N$, $\mathcal{V}=\mathcal{V}^1\cup \mathcal{V}^2\cup .... \mathcal{V}^N$ and $\mathcal{R}=\mathcal{R}^1\cup \mathcal{R}^2\cup .... \mathcal{R}^N$ as user set, item set and interaction set of all the domains, respectively.

\textbf{Single-domain Sequential Recommendation.} Given $\mathcal{U}^n, \mathcal{V}^n$ and $\mathcal{R}^n$ in domain $\mathcal{D}^n$, we can organize the user $u$'s interaction history from domain  $\mathcal{D}^n$ as a sequence $s_u^n=(v_1^n, v_2^n, ..., v_L^n)$ following their chronological order, where $v_i^n \in \mathcal{V}^n$ and $L$ is the length of sequence. Hence, single-domain sequential recommendation aims to predict the probability of the next item $v_{L+1}^n$  based on $s_u^n$.

\textbf{Multi-domain Sequential Recommendation.} Given $\mathcal{U}, \mathcal{V}$ and $\mathcal{R}$ from all the domains. For each user $u$ in $\mathcal{U}$, we can collect the user's interaction history from all domains and organize a mixed sequence $s_u=(v_1, v_2, ..., v_{L'})$ chronologically, where $(v_1,..., v_{L'})$ may come from different domains. Then, given a target domain $\mathcal{D}^T$, multi-domain sequential recommendation aims to predict the probability of the next item $v_{L'+1}^T \in \mathcal{V}^T$ based on the mixed sequence.

\subsection{\baby}

\subsubsection{Transformer}
Due to the excellent parallelizability and capacity, the Transformer ~\cite{Transformer} has become the most popular basic architecture for developing various large language models. We will give a brief introduction about the Transformer at first.
\begin{itemize}
    \item \textbf{Embedding Layer.}
    Embedding layer is the first component of the Transformer, which contains a word embedding table $E$ and position embedding table $P$. Given a tokenized word token sequence $T=(t_1, t_2, ..., t_C)$, each word token $t_i$ is mapped to a unique vector $e_i$ in the word embedding table. These embeddings capture semantic and contextual information about the words, enabling the model to understand the meaning and relationships between them. Additionally, there is no inherent notion of token order or position information in the input sequence, as the Transformer processes the sequence in parallel. The position embedding table provides a set of vectors $(p_1, p_2, ..., p_C)$ that represent the position of each token in the input sequence and are added to the word embeddings as:
    \begin{equation}
        F_0=[e_1+p_1, e_2+p_2, ..., e_C+p_C]
    \end{equation}
    where $C$ represents the token sequence length and $F^0$ refers to the input for the Transformer model. By combining the word embeddings with the position embeddings, the Transformer model can effectively encode both the meaning of individual word token and the sequential patterns of the entire input sequence.
    \item \textbf{Self-Attention Block Layer.} The self-attention block layer receives a sequence of representations from the previous layer as input and generates a sequence of more contextual representations as output. And the first layer takes the output of the Embedding Layer as its input. Each self-attention block layer is composed of a self-attention module and a point-wise feed-forward network. Self-attention module allows the Transformer model to capture dependencies of tokens and produces more accurate and context-aware representations. The definition of the $l-th$ layer scaled dot-product self-attention is as follows:
    \begin{equation}
        S_{l} = Self-Attention_l(F_{l-1}) = softmax(\frac{Q_lK_l^T}{\sqrt{d}})V_l
    \end{equation}
    \begin{equation}
        Q_l, K_l, V_l = W_l^QF_{l-1}, W_l^KF_{l-1}, W_l^VF_{l-1}
    \end{equation}
    where $F_{l-1}$ is the input sequence representation, $d$ is latent dimensionality and $W^Q_l, W^K_l, W^V_l$ are the projection matrices. Furthermore, the output of self-attention is passed through a residual connection and LayerNorm before being input to the Position-wise Feed-Forward Network(PPN) with another residual connection and LayerNorm:
    \begin{equation}
        F_l = LayerNorm(PPN(S_l)+S_l)
    \end{equation}
    \begin{equation}
        PPN(S_l) = LayerNorm(Relu(S_lW^0_l)W^1_l+F_l)
    \end{equation}
    where $W^0_l, W^1_l$ are the projection matrices and $Relu$ is the activation function. By stacking more self-attention blocks, the Transformer gains stronger modeling capabilities and more contextual information. Hence, for a Transformer with $L$ self-attention block layers, the output of the $L$-th self-attention block is the contextualized output of the Transformer.
    Note that different language models may be based on different variants of the Transformer, and the treatment mentioned above is for the vanilla Transformer.
    
\end{itemize}

\subsubsection{Backbones}
As aforementioned, we use three types of standard networks of LLM as our backbone for investigation.

\begin{itemize}

    \item \textbf{Encoder-only.} Encoder-only architecture utilizes bi-directional attention (ref. the bottom left part in Figure ~\ref{fig:framework}) for context information modeling and mask token prediction. We choose BERT~\cite{BERT} as the representative.
    \item \textbf{Decoder-only.} Different from Encoder-only architectures, Decoder-only utilizes a uni-directional (ref. the bottom right part in Figure ~\ref{fig:framework}) attention for the next token generation. Such that, only the previous information could be obtained via the self-attention to generate the next token. This paradigm holds a dominant position in the design of autoregressive LLM nowadays. We, thereby, apply Open Pre-trained Transformers (OPT)~\cite{Opt}, a suite of decoder-only pre-trained transformers ranging from 125M to 175B parameters, as another backbone of our \baby.
    \item\textbf{Encoder-Decoder.} The Encoder-Decoder architecture is another prevalent design approach employed in LLM, which contains both an encoder (i.e., responsible for transforming input data into a fixed-length representation) and a decoder (i.e., tasked with generating the output sequence based on the encoded representation). We choose FLAN-T5~\cite{FLAN-T5}, a variant of T5 that is fine-tuned with prompting in a mixture of tasks, as our backbone. It should be noted that we use language models for representation learning. Hence, we only use the encoder of FLAN-T5.
\end{itemize}

\subsubsection{Model Inputs}
In our paper, we use the item title $t_v$ instead of ID as the input, where $t_v=(w_{v}^1,...,w_{v}^K)$ and $w_v^k$ is the token after tokenization.
Following the common practices in NLP, we add a special token $[CLS]$ at the beginning of the text sequence for bi-directional Transformer and append the $</s>$ at the end of the sequence with regard to the uni-directional Transformer. 
We can also call these two architectures as non-autoregressive (NAR) and autoregressive (AR) respectively.
Consequently, each item can be represented as below:

\begin{itemize}
    \item \textbf{Item Input}
\end{itemize}
\begin{equation}
Input(v)=
\begin{cases}
[[CLS],w_{v}^1, ..., w_{v}^K], & \text{NAR}; \\
[w_{v}^1, ..., w_{v}^K,</s>], & \text{AR}; 
\end{cases}
\label{eq:item}
\end{equation}

As for the user representation, we concatenate each item representation yield by Equation~\ref{eq:item} as a sentence for user representation, which can be formalized as:
\begin{itemize}
    \item \textbf{User Input}
\end{itemize}
\begin{equation}
Input(u)=
\begin{cases}
[[CLS],Input_{w/o}(v_1), ..., Input_{w/o}(v_{L'})]; \\
[Input_{w/o}(v_1), ..., Input_{w/o}(v_{L'}),</s>]; 
\end{cases}
\label{eq:user}
\end{equation}

where $Input_{w/o}(v)$ means removing the $[CLS]$ or $</s>$ token from $Input(v)$.

\subsubsection{Representation and Prediction}
Both user and item input will be further fed into the backbone for sequence modeling and representation generation. To be specific, for the models with NAR Transformer architecture, we obtain the correspondent hidden vector of $[CLS]$ in the last layer as the user or item representation. As for the AR Transformer architecture, the derived hidden representation of the last layer with regard to $</s>$ is utilized, which can be formulated as follows: 

\begin{equation}
    h_u = backbone(Input(u))
    \label{eq:inner_u}
\end{equation}
\begin{equation}
    h_v = backbone(Input(v))
    \label{eq:inner_v}
\end{equation}

where $h_u \in \mathbb{R}^{1 \times d}$ and $h_v \in \mathbb{R}^{1\times d}$ are the representation of user and item, respectively.

We predict the next item based on the score between a user's representation $h_u$ and candidate item representation $h_v$ yielded by Equation~\ref{eq:inner_u} and~\ref{eq:inner_v}. Formally, we calculate the inner product between them as follows:
\begin{equation}
    \Tilde{y}_{u,v}=h_uh_v^T
\end{equation}
\subsubsection{Optimization}
For each target item, we randomly sample $S$ negative instances from the same domain as the hard negatives and adopt the cross-entropy loss as the objective function for \baby optimization:
\begin{equation}
    \ell = \sum_{j=1}^{B}-log\frac{exp(\Tilde{y}_{u_j,v_j})}{exp(\Tilde{y}_{u_j,v_j})+\sum_{v_{j'} \in \mathcal{S}}exp(\Tilde{y}_{u_j,v_{j'}})}
\end{equation}
where $B$ is the batch size, $u_j$ and $v_j$ represent the $j$-th user and ground truth item in the batch respectively.
\section{Experiments}
\label{sec:exp}
To demonstrate the effectiveness of our methods and answer the research question in Section~\ref{sec:intro}, we conduct extensive experiments on five public datasets, comparing the performance between our \baby with three types of backbones and $8$ representative baselines. 
\subsection{Experimental Setup}
\label{Experimental Setup}
We first present the statistics of datasets in brief. Then, the advanced baselines, evaluation metrics, and parameter settings will also be given.

\subsubsection{Datasets}
Considering the need for multi-domain user interactions and high-quality textual features, we align with previous methods ~\cite{UniSRec, MIFN, SCDR_2, Recformer} by adopting the Amazon dataset to validate \baby's effectiveness.
We select five sub-category datasets from the Amazon reviews\footnote{\url{https://nijianmo.github.io/amazon/index.html}} dataset, namely "Prime Pantry", "Industrial and Scientific", "Musical Instruments", "Arts, Crafts and Sewing", and "Office Products", each of which represents a different domain. 
Firstly, we remove the items without title information in meta-data as the title will be used for item identification, symbolization, and modeling.
Afterward, we extract historical interaction behaviors from these five domains for each user and order them chronologically for sequence construction.
Then, following~\cite{UniSRec, SASRec, S3-ReC}, a 5-core strategy is applied for item and user filtering, i.e., each user is required to contain at least five interaction records from multi-domains, and each item should appear no less than five times. 
We leverage a leave-one-out strategy for training, validation, and test dataset split. More concretely, given a user sequence, we use the latest interaction for testing, the penultimate interaction for validation, and the remaining interactions for training.
We denote the pre-processed multi-domain dataset as SPAIO, and the statistical information is presented in Table~\ref{tab:statisitics_1}-~\ref{tab:statisitics_3} respectively. We can observe that over 80\% of users have interactions in more than one domain, and there is also a significant variance in the number of interactions across different domains.

{
\begin{table}[t]
\setlength\tabcolsep{3pt}
\centering
\caption{The statistics of the datasets.}
\label{tab:statisitics_1}
\begin{tabular}{c|rrrrr}
\bottomrule
Dataset       & \#Users       & \#Items     & \#Interactions   & \#Training   & \#Testing   \\ \hline\hline 
Scientific    & 142,139     & 14,586    & 329,123        & 244,960    & 43,365         \\ 
Pantry        & 55,890      & 6,224     & 208,791        & 162,762    & 22,412         \\ 
Instruments   & 88,249      & 14,626    & 348,513        & 274,760    & 36,129         \\ 
Arts          & 184,207     & 32,196    & 723,017        & 568,600    & 76,035         \\ 
Office        & 259,687     & 37,380    & 1,154,924      & 893,014    & 132,195         \\  
SPIAO         & 310,136     & 105,012   & 2,764,368      & 2,144,096  & 310,136         \\
\toprule
\end{tabular}
\end{table}
}

{
\begin{table}[t]
\setlength\tabcolsep{3pt}
\centering
\caption{User overlap between different domains.}
\label{tab:statisitics_2}
\begin{tabular}{c|rrrrr}
\bottomrule
Dataset         & Scientific    & Pantry    & Instruments   & Arts      & Office        \\  \hline
Scientific      & -             & -         & -             & -         & -             \\
Pantry          & 26,863        & -         & -             & -         & -             \\
Instruments     & 39,806        & 10,812    & -             & -         & -             \\
Arts            & 81,855        & 30,807    & 38,353        & -         & -             \\
Office          & 124,621       & 47,206    & 65,487        & 155,995   & -             \\
\toprule
\end{tabular}
\end{table}
}

{
\begin{table}[t]
\centering
\caption{The number of users that have records across different domains.}
\label{tab:statisitics_3}
\begin{tabular}{c|cccccc}
\bottomrule
Number of Domains & 1      & 2        & 3      & 4      & 5      & ALL       \\  \hline
Number of Users   & 53,117 & 129,352  & 95,719 & 28,546 & 3,402  & 310,136    \\
\toprule
\end{tabular}
\end{table}
}

{\centering
\begin{table*}[t] 
\caption{Performance comparison of different methods, with \% omitted. The best results are highlighted in bold, and the best performance of baselines is underlined. It should be noted that NDCG@N is equal to Recall@N when N=1, so we omit the result of NDCG@1 for simplicity. For different LLMs, we report results of similar model sizes (around 110M).}

\label{tab:overall}

\resizebox{\textwidth}{!}{
\setlength{\tabcolsep}{2pt} 
\begin{tabular}{cc|cccc|ccc|cc|ccc}
\bottomrule
\multirow{2}{*}{Dataset}     &  \multirow{2}{*}{Metrics} & \multicolumn{4}{c}{Single Domain}   & \multicolumn{3}{c}{Multi Domain} & \multicolumn{2}{c}{Text-Based} & \multicolumn{3}{c}{\baby}  \\
 &     &  \small{GRU4Rec}  &  \small{SASRec} &\small{FEARec} &  \small{LightGCN}   & \small{ADI}    & \small{BiTGCF+ }    & \small{MAMDR}  &\small{UniSRec$_t$} &\small{MoRec}  &\small{BERT}    & \small{OPT}    &\small{FLAN-T5}  \\ \hline\hline
\multirow{3}{*}{Scientific}  & Recall@1  & 5.31     & 6.22      & \underline{6.85}      & 4.38      & 4.03  & 4.53      & 6.35 & 5.46 & 5.68 & \textbf{7.81}      & 7.36      & 7.66          \\  
                             & Recall@10 & 18.70    & 19.16     & 19.45     &16.90      & 15.96 & 17.43     & \underline{20.85}& 16.62 & 21.40 &\textbf{25.61}     & 24.26     & 25.56         \\  
                             & NDCG@10   & 11.09    & 11.89     & 12.38     &9.85      & 9.17  & 10.15     & \underline{12.71} & 10.44 & 12.54 &\textbf{15.63}     & 14.79     & 15.58         \\ \hline 
\multirow{3}{*}{Pantry}      & Recall@1  & 2.81     & 3.54      & \underline{4.29}      &2.30      & 1.99  & 2.30      & 3.77  & 3.72 & 3.42 &\textbf{5.36}      & 4.37      & 4.83          \\ 
                             & Recall@10 & 13.68    & 15.18     & 15.84     &10.70      & 10.39 & 11.34     & \underline{16.53} & 15.16 & 16.41 & \textbf{20.42}     & 18.92     & 20.00         \\ 
                             & NDCG@10   & 11.09    & 8.71      & 9.37      &5.91      & 5.54  & 6.16      & \underline{9.40}  & 8.85 & 9.09 & \textbf{12.11}     & 10.83     & 11.54         \\ \hline 
\multirow{3}{*}{Instruments} & Recall@1  & 7.13     & 7.86      & \underline{8.62}      &4.73      & 5.49  & 5.16      & 8.31  & 6.36 & 6.96 & \textbf{8.92}      & 8.41      & 8.74          \\ 
                             & Recall@10 & 22.08    & 23.30     & 24.11      &20.52     & 20.63 & 21.45     & \underline{26.12} & 20.36 & 25.09 & \textbf{30.57}     & 28.91     & 30.51         \\ 
                             & NDCG@10   & 13.63    & 14.63     & 9.37     &11.58     & 12.02 & 12.23     & \underline{16.14} & 12.63 & 14.84 & \textbf{18.41}     & 17.45     & 18.28         \\ \hline
\multirow{3}{*}{Arts}        & Recall@1  & 8.48     & 10.77     & \underline{11.80}     &5.54      & 6.63  & 5.89      & 11.12 & 8.81 & 9.13 & \textbf{13.46}     & 12.48     & 12.77         \\  
                             & Recall@10 & 25.31    & 28.25     & 28.35     &22.73     & 23.09 & 23.83     & \underline{30.73} & 24.70 & 30.66 & 36.58              & 35.01     & \textbf{36.78}         \\ 
                             & NDCG@10   & 15.89    & 18.61     & 19.19     &12.96     & 13.76 & 13.66     & \underline{19.91} & 15.86 & 18.63 & \textbf{23.90}     & 22.63     & 23.59         \\ \hline
\multirow{3}{*}{Office}      & Recall@1  & 8.58     & 10.44     & \underline{11.15}     &5.27      & 7.50  & 5.66      & 10.91 & 9.35 & 8.68 & \textbf{12.75}     & 12.33     & 12.40         \\ 
                             & Recall@10 & 23.64    & 25.14     & 24.74     &20.27     & 23.30 & 23.83     & \underline{27.41} & 23.64 & 26.58 & 32.58              & 32.02     & \textbf{32.74}         \\ 
                             & NDCG@10   & 15.16    & 16.94     & 17.13     &11.80     & 14.40 & 13.66     & \underline{18.18} & 15.75 & 16.54 & \textbf{21.59}     & 21.16     & 21.50         \\ \hline
\multirow{3}{*}{SPIAO} 
                             & Recall@1  & 7.50     & 9.13      & \underline{9.92}      &4.93      & 6.17  & 5.26      & 9.50  & 7.93 & 7.79 &\textbf{11.25}     & 10.64     & 10.85         \\ 
                             & Recall@10 & 22.46    & 24.13     & 24.17      &19.75     & 20.98 & 20.44     & \underline{26.37} & 21.93 & 25.95 & 31.47              & 30.36     & \textbf{31.55}         \\ 
                             & NDCG@10   & 14.03    & 15.78     & 16.21     &11.36     & 12.59 & 11.86     & \underline{16.97} & 14.18 & 15.76 &\textbf{20.27}     & 19.45     & 20.09         \\ \toprule
\end{tabular}}

\end{table*}
}

\subsubsection{Comparing Methods}
\label{Baselines}
Here, we compare the performance of our \baby against six baselines and state-of-the-art alternatives from three categories including single-domain methods, multi-domain methods, and text-based methods for performance evaluation. For single-domain and text-based methods, we regard the mixed dataset SPIAO as a single domain. 

\textbf{Single-domain methods:}
\begin{itemize}
\item \textbf{GRU4Rec}~\cite{GRU4Rec} utilizes a gate recurrent unit model for sequence recommendation.
\item \textbf{LightGCN}~\cite{LightGCN} is a graph neural network (GNN) based model. It captures the high-order connectivity information by stacking the GNN layer and simplifies the design in the feature propagation component by removing the non-linear activation and the transformation matrices in the GNN.
\item \textbf{SASRec}~\cite{SASRec} employs a uni-directional self-attention mechanism for sequence modeling and next item prediction.
\item \textbf{FEARec} ~\cite{FEARec} enhances sequential recommendation by integrating frequency domain insights to capture long-range dependencies and inherent periodicities.

\end{itemize} 

\textbf{Multi-domain methods:}
\begin{itemize}
\item \textbf{ADI}~\cite{ADI} models multi-domain recommendation as multi-task learning, which includes both domain-specific expert networks and domain-shared expert networks. A  self-training strategy is used to capture label-level connections between domains.
\item \textbf{BiTGCF}~\cite{BiTGCF} is a dual-target cross-domain model based on a graph collaborative filtering network. It designs a feature propagation layer for different domain information transfers. We extend it to be applicable for multi-domain recommendation denoted as "BiTGCF+".
\item \textbf{MAMDR}~\cite{MAMDR} MAMDR is a model-agnostic multi-domain recommendation learning framework, which proposes the Domain Negotiation strategy to alleviate gradient conflicts during multi-domain training. Additionally, Domain Regularization is applied to enhance the generalizability to the other domains. In our work, we select SASRec as the backbone for multi-domain recommendation since MAMDR is a model-agnostic method.
\end{itemize}

\textbf{Text-Based Methods:}
\begin{itemize}
    \item \textbf{UniSRec}~\cite{UniSRec} utilizes item representations derived from a pre-trained language model and adjusts to a new domain through an adaptor enhanced by Mixture of Experts (MoE). We use its inductive setting UniSRec$_t$ which does not rely on the item ID.
    \item \textbf{MoRec}~\cite{MoRec} investigates the feasibility of using modality representations instead of ID embeddings, demonstrating the potential of modality representation. In this paper, we use an unfrozen BERT-base pre-trained model as the modality encoder.
\end{itemize}

\subsubsection{Evaluation Metrics}
\label{Evaluation Metrics}
\label{Implementation Details}
We use Recall@N and NDCG@N for recommendation performance evaluation, where $N=1$ and $10$.
Compared with Recall,  NDCG further takes the position of retrieved positive item into account and assigns higher score when the positive items are ranked higher in the retrieved results. For all the datasets, we use negative sampling for evaluation, i.e., for each target item we randomly sample $1,000$ instances from the domain of the target item as negatives.

\subsubsection{Implementation Details}
All methods are implemented using Pytorch with an Adam~\cite{Adam} optimizer. 
Due to the large size of the item set and to enable more appropriate comparisons, we choose sampled cross-entropy loss for all the methods. The negative sample count of ID model and UniSRec$_t$ is set to 64, and the MoRec and \baby, due to their high memory usage, the negative sample count is set to 10.
The language models mentioned in our \baby are from Huggingface\footnote{\url{https://huggingface.co/docs/transformers/model_doc/flan-t5}}\footnote{\url{https://huggingface.co/docs/transformers/main/model_doc/opt}}\footnote{\url{https://huggingface.co/docs/transformers/main/model_doc/bert}}. We set the maximum length of the interaction sequence to 10 and truncate the length of item title to $40$. The learning rate is $5e-5$ for the OPT, BERT and their variants, and $3e-4$ for FLAN-T5 and its variants.
The batch size is set to 96 for all models.
We check the validation performance every $3,000$ steps and adopt the early-stop strategy for our model training, i.e., the training process will be terminated when the performance of the validation dataset is not improved after ten consecutive rounds.

\subsection{Performance Under Multi-domain Settings (Q1)}

\subsubsection{Overall Comparison}
\label{sec:perf}
In Table ~\ref{tab:overall}, we report the results of baselines and our \baby with different backbones in five domains as well as the mixed SPIAO dataset. The results of SPIAO are the weighted average of the proportions of five domains in the test set.
From the experimental results, we could draw the following conclusions: 
1) As an extension of LightGCN in multi-domains, BiTCGF+ outperforms LightGCN, which demonstrates the effectiveness of multi-domain information for recommendation. However, BiTGCF+ is inferior to single-domain sequential methods (e.g., SASRec). It illustrates that it is non-trivial to apply graph-based models for sequential recommendation as they could not capture the long-term dependency in sequences effectively. 
2) Compared to SASRec, MAMDR acquires stable improvements in all domains demonstrating the value of multi-domain data. However, ID-based models require complex designs for the model or the optimization process.
3)The text-based baseline methods, which utilize a pre-trained model to encode items, have not yielded satisfactory outcomes. We believe the reason for this is that both methods first use a pre-trained language model to encode the text into a single representation, and then use a sequence model, making it difficult to directly capture the word-level associations between item texts. They merely treat the text as an additional feature, which is not optimal for multi-domain scenarios.  Yet, LLM-Rec is also a purely textual approach, and it employs language models directly for modeling user representations; the items interacted with by the same user can achieve fine-grained interactions at the textual level and achieve better performance.
4) Besides, different variants of \baby achieve the optimal results, presenting the superiority of language models for the multi-domain recommendation, different variants of LLMRec achieved, on average, a relative improvement of over 19\% on the SPIAO dataset compared to the optimal baseline.
Moreover, we find that the results of \baby will fluctuate slightly under different model structures, although their model sizes are close. 

In the following sections, we will analyze the impact of cross-domain data, language model size, and fine-tuning methods in-depth.

\begin{figure}[t]
    \centering{
    \subfigure{\includegraphics[width=1.0\textwidth]{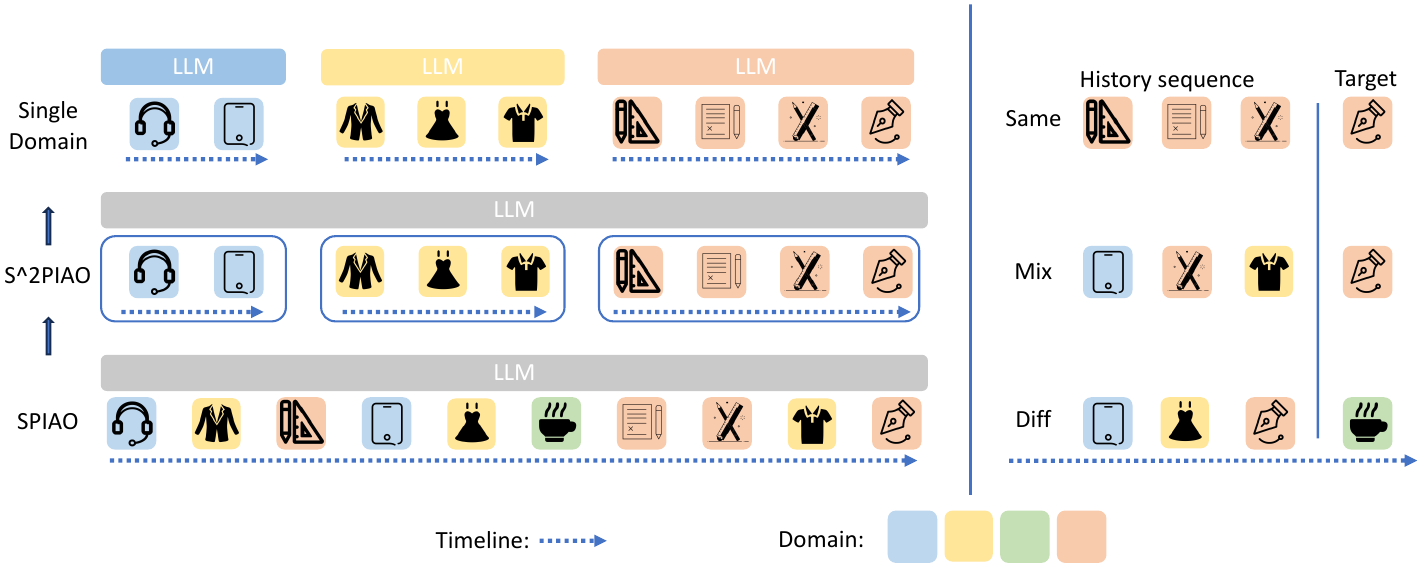}}
    \caption{Two different cross-domain settings. The left half describes the partitioning of different training sets and the right half describes the partitioning methods for different testing sets. It is worth noting that in the left part item represented by a green background is ignored in both Single Domain and S$2$PIAO settings due to only one occurrence in the sequence. }
    \label{fig:data_setting}
    }
\end{figure}

\begin{figure}[htbp]
    \centering{
    \subfigure{\includegraphics[width=0.3\textwidth]{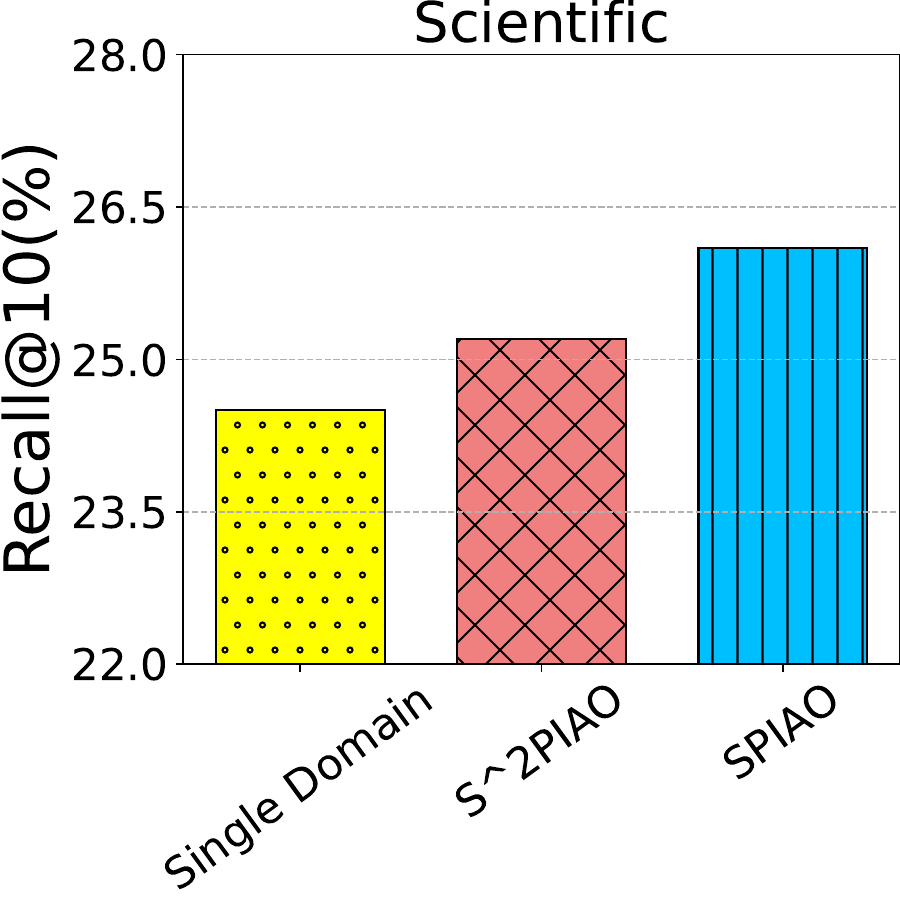}}
    \subfigure{\includegraphics[width=0.3\textwidth]{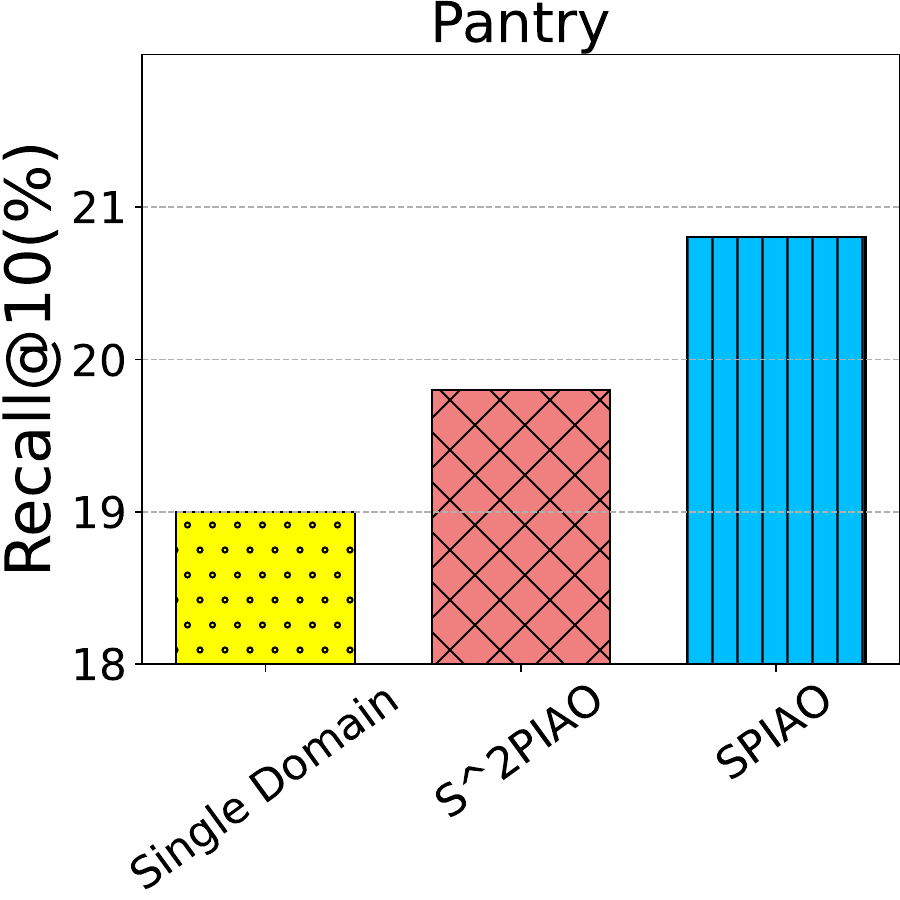}}
    \\
    \subfigure{\includegraphics[width=0.3\textwidth]{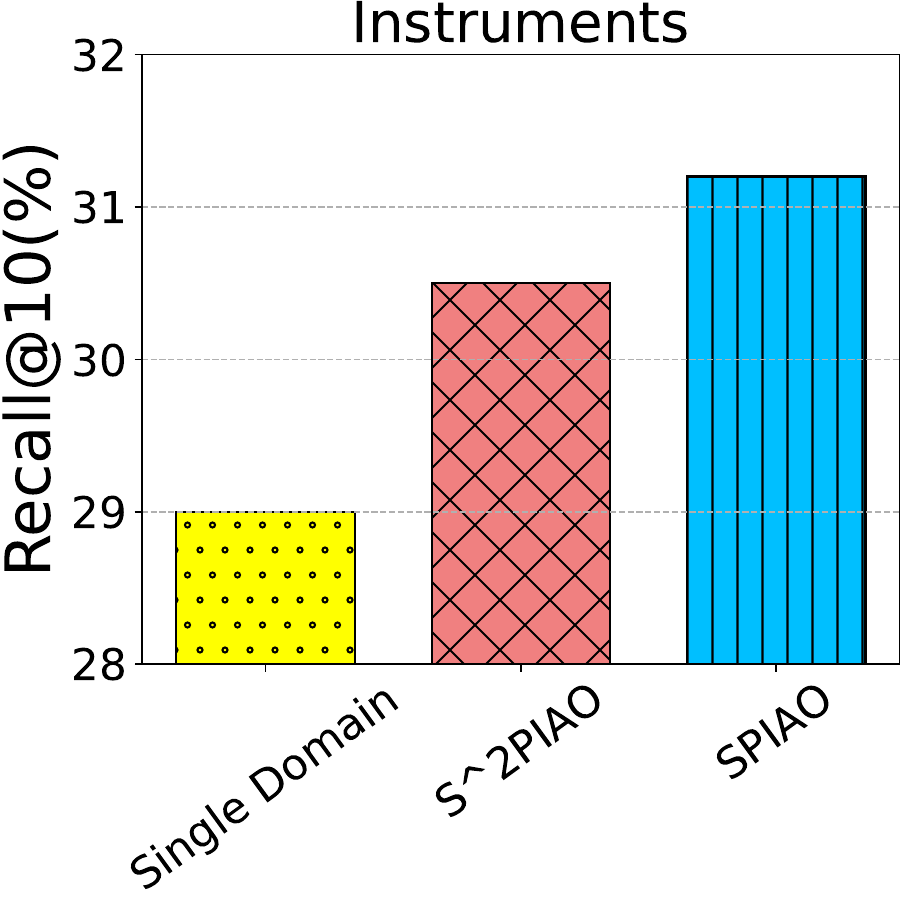}}
    \subfigure{\includegraphics[width=0.3\textwidth]{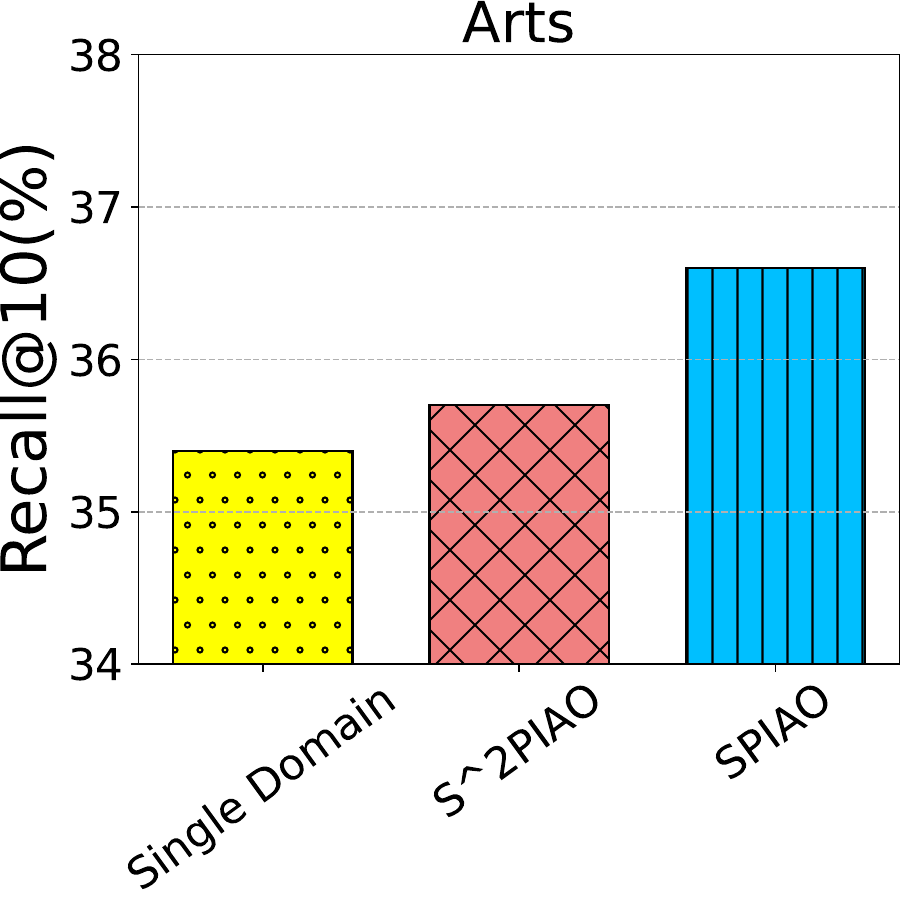}}
    }
    \caption{Performance under different cross-domain data settings. The pre-trained language model is BERT-110M.}
    \label{fig:ablation_1}
\end{figure}

\begin{figure}[htbp]
    \centering{
    \subfigure{\includegraphics[width=0.45\textwidth]{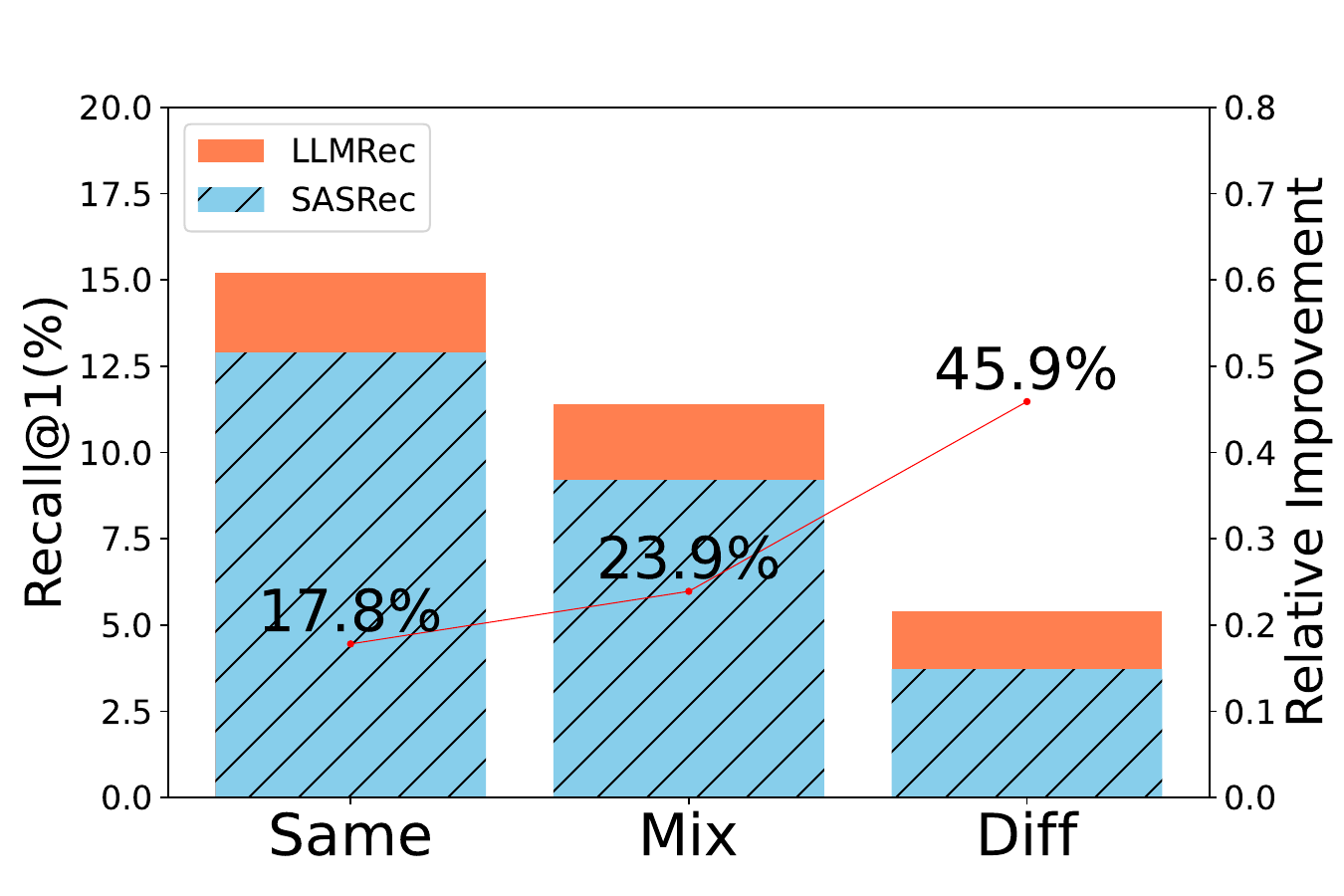}}
    \subfigure{\includegraphics[width=0.45\textwidth]{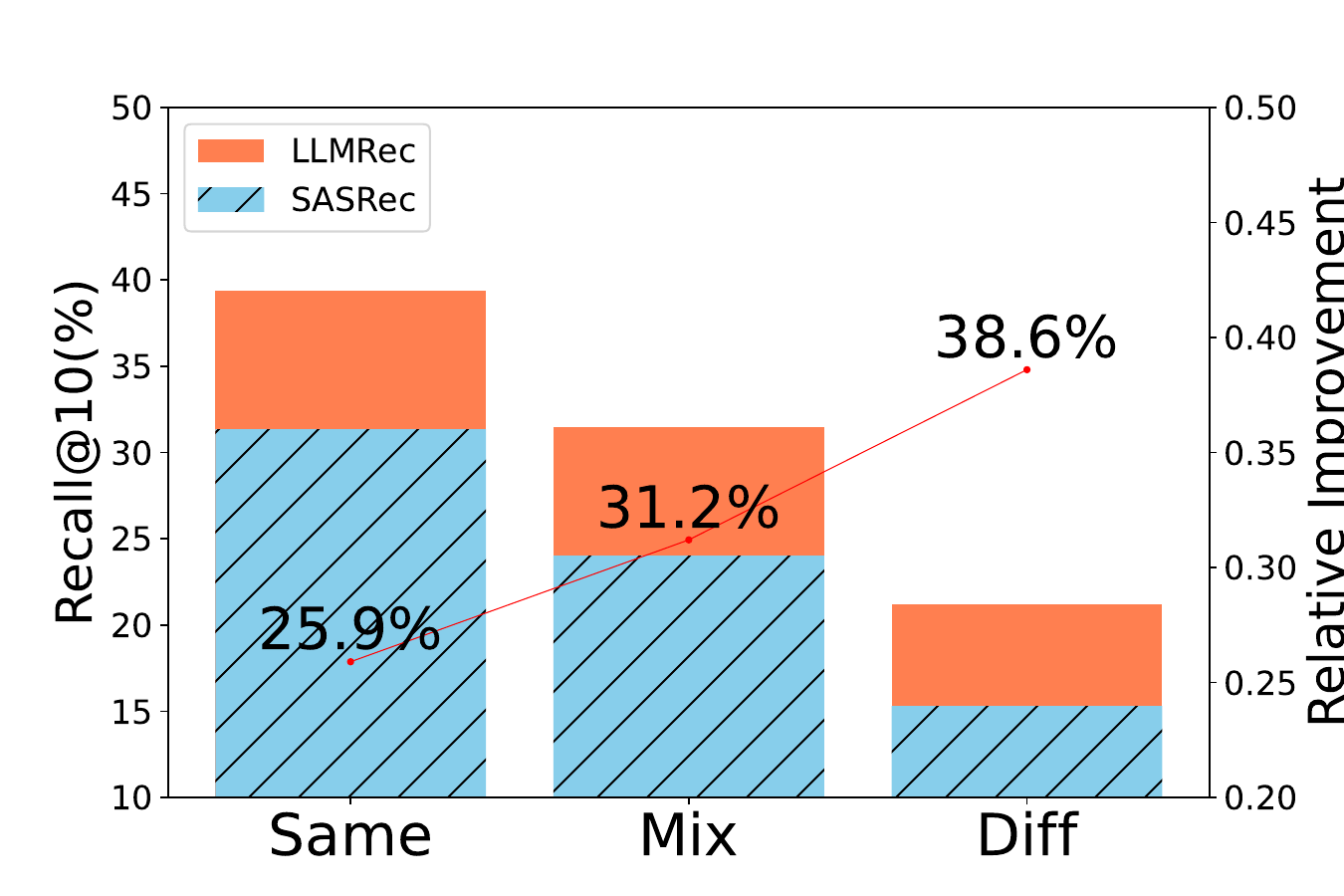}}
    }
    \caption{The relative improvement under different sequence settings, the pre-trained language model is BERT-110M (\textit{Best viewed in color})}
    \label{fig:ablation_2}
\end{figure}

\subsubsection{The Impact of Cross-domain Data}
\label{sec:impact_data}

\paragraph{Without Cross-Domain Data.}To verify the benefits of cross-domain interactions, we further analyze the performance changes with and without the cross-domain records during training. To be specific, we first organize the user interaction sequence for each domain (e.g., Scientific, Pantry, Instruments, etc.) and then mix them together as the multi-domain dataset denoted as S$^2$PIAO, and the data settings are visualized in the left part of Figure ~\ref{fig:data_setting}. It is noteworthy that in S$^2$PIAO each user will contain several interaction sequences, each of which corresponds to a specific domain. Thus, for each sequence, there is no cross-domain information sharing, while for SPIAO, each individual will only obtain a single mixed sequence formed by sorting all domains' interaction behaviors along the timeline. For in-depth analysis, we denote the S$^2$PIAO mixing method as the domain-oriented mix strategy and denote SPIAO as the user-oriented mix strategy. Then, we retrain our model with single-domain datasets (e.g., Scientific, Pantry, Instruments, and so on and so forth.) and multi-domain dataset (S$^2$PIAO) and SPIAO, respectively, and compare their performance under the same test dataset. The result is shown in Figure~\ref{fig:ablation_1}. 

Observing Figure~\ref{fig:ablation_1}, we can find that the SPIAO achieves the best performance and single-domain performs the worst, indicating that a simplified mixing solution (i.e., domain-oriented mix strategy) bring lots of performance gain for recommendation. Moreover, it is obvious that the user-oriented finer grain mixing can further improve the model performance. We believe this improvement comes from the world knowledge of pre-trained LLM. And richer and more detailed description of the user sequence will reveal more information.

\paragraph{Performance Improvement under Different Sequence Settings.}To further investigate the performance of our \baby, we divide the test set into three categories based on the relationship of the target item domain and sequence item domain and compare the performance between \baby against SASRec under these three settings as below, it should be noted the model is still trained on the mixed SPIAO dataset, but only the test set is partitioned for evaluation. The data settings are visualized in the right half of Figure ~\ref{fig:data_setting}:
\begin{itemize}
    \item \textbf{Same}: the target item and the items in the sequence are all from the same domain. 
    \item \textbf{Mix}: a part of the items in the sequence is from the target item domain, while the left is from other domains.
    \item \textbf{Diff}: the domains of all the items in the sequence are orthogonal to the target item's domain.
\end{itemize}

Figure~\ref{fig:ablation_2} shows the results, we can find that the relative improvement of  \textbf{Diff} setting is superior to the \textbf{Mix} significantly, which in turn is superior to the \textbf{Same}. 
This observation supports our assumption that as a new alternative to the recommended solution, LLM could achieve competitive results compared with conventional ID-based sequential modeling methods. Furthermore, the significant improvement of \baby on cross-domain data further verifies our idea that world knowledge embedded in a pre-trained language model can facilitate multi-domain knowledge transfer. It is worth noting that in Figure 6, there is no overlap of users across different settings, so the absolute performance between different settings are not comparable. We aim to verify that language models facilitate knowledge transfer across domains by comparing the relative performance improvements achieved in different user groups.

\subsubsection{Case Study}
\label{sec:case_study}
{
\begin{table}[htbp]
\setlength\tabcolsep{5pt}
\centering
\caption{Case study under ‘Diff’ setting that the target domain is different from all the interacted domains. The same interests in each case are highlighted by the same color. "GT" represents the ground truth item, and "\checkmark" means the ground truth item in the recommendation list.}
\label{tab:case_study}
\renewcommand\cellalign{tl}
\begin{tabular}{c|c|r|l}
\bottomrule
Case 1  & \multicolumn{2}{l|}{Interaction Sequence} & Category \\ \hline
        & \multicolumn{2}{l|}{\makecell{1. \textcolor{blue}{adc adscope 615 platinum professional clinician stethoscope with }\\\textcolor{blue}{tunable afd technology}}}  & Scientific\ \ \ \ \ \   \\      
        & \multicolumn{2}{l|}{2. \textcolor{red}{bic round stic xtra life ballpoint pen, medium point, black.}} & Office \\  
        & \multicolumn{2}{l|}{3. \textcolor{red}{bic round stic xtra life ballpoint pen, medium point, red.}} &  Office \\  
        & \multicolumn{2}{l|}{4. \textcolor{red}{expo low odor chisel point dry erase marker pack.} } &  Office \\ 
        & \multicolumn{2}{l|}{5. \textcolor{red}{bicvlgb11be - bic velocity easy-glide system ballpoint pen.} } &  Office \\ 
        & \multicolumn{2}{l|}{6. \textcolor{blue}{3m littmann stethoscope spare parts kit, cardiology iii.} } &  Scientific \\ 
        \hline
GT   & \multicolumn{2}{l|}{ever ready titanium bonded bandage shears.} & \multirow[t]{3}{*}{Arts} \\ \cline{1-3}
        & \multicolumn{2}{l|}{\baby List} & \\ 
        & \multicolumn{2}{l|}{1. \textcolor{red}{pentel hi - polymer block eraser}}  &  \\ 
        & \multicolumn{2}{l|}{2. sparco spr01796 heavy-duty adjustable.} &  \\  
        & \multicolumn{2}{l|}{3. \textcolor{blue}{ever ready titanium bonded bandage shears.}} &  \checkmark \\  
        & \multicolumn{2}{l|}{4. \textcolor{red}{loew-cornell 245b brush set, multi color.}} &  \\ 
        & \multicolumn{2}{l|}{5. evergreen art supply super scissors.} &  \\  
        \cline{1-3}
        & \multicolumn{2}{l|}{IDRec List} & \\ 
        & \multicolumn{2}{l|}{1. cindus tissue wrap}  &  \\ 
        & \multicolumn{2}{l|}{2. testors 9111xt hobby supplies paint kit.} &  \\  
        & \multicolumn{2}{l|}{3. brother se400 combination computerized sewing.} &   \\  
        & \multicolumn{2}{l|}{4. \textcolor{red}{loew-cornell 245b brush set, multi color.}} &  \\ 
        & \multicolumn{2}{l|}{2. sparco spr01796 heavy-duty adjustable.} &  \\  
        \hline
\toprule
\end{tabular}

\begin{tabular}{c|c|r|l}
\bottomrule
Case 2  & \multicolumn{2}{l|}{Interaction Sequence} & Category \\\hline
        & \multicolumn{2}{l|}{\makecell{1. \textcolor{blue}{blue snowball usb microphone(brushed aluminum)}}}  & Instruments \\      
        & \multicolumn{2}{l|}{2. {bbloop stamp; rectangular. laser engraved. red.}} & Office \\  
        & \multicolumn{2}{l|}{3. {bic round stic xtra life ballpoint pen, medium point, red.}} &  Office \\  
        & \multicolumn{2}{l|}{4. fisher space pen pressurized refill, medium point, black ink(spr4).}  &  Instruments \\ 
        & \multicolumn{2}{l|}{5. \textcolor{blue}{on-stage ms7701b tripod microphone boom stand.}} &  Office \\ 
        & \multicolumn{2}{l|}{6. fisher space pen men's bullet space pen with clip. } &  Scientific \\ 
        \hline
GT   & \multicolumn{2}{l|}{audioengine d1 24-bit digital-to-analog converter.} & \multirow[t]{3}{*}{Arts} \\ \cline{1-3}
        & \multicolumn{2}{l|}{\baby List} & \\ 
        & \multicolumn{2}{l|}{1. arctic silver 5as5-3.5g thermal paste}  &  \\ 
        & \multicolumn{2}{l|}{2. niteize lsbm-01-2r3s-biner microlock stainless steel, 2-pack, black.} &  \\  
        & \multicolumn{2}{l|}{3. gorilla 6100101 tape handy roll, 1 - pack, black.} &   \\  
        & \multicolumn{2}{l|}{4. \textcolor{blue}{audioengine d1 24-bit digital-to-analog converter.}} & \checkmark  \\ 
        & \multicolumn{2}{l|}{5. se 82331tf 30-piece set of titanium-coated diamond burrs.} &  \\  
        \cline{1-3}
        & \multicolumn{2}{l|}{IDRec List} & \\ 
        & \multicolumn{2}{l|}{1. niteize lsbm-01-2r3s-biner microlock stainless steel, 2-pack, black.}  &  \\ 
        & \multicolumn{2}{l|}{2. gorilla 6100101 tape handy roll, 1 - pack, black.} &  \\  
        & \multicolumn{2}{l|}{3. arctic silver 5as5-3.5g thermal paste.} &   \\  
        & \multicolumn{2}{l|}{4. gorilla super glue gel, 15 gram, clear} &  \\ 
        & \multicolumn{2}{l|}{2. crc 05002 freeze-off super penetrant.} &  \\  
        \hline
\toprule
\end{tabular}
\end{table}
}

In the previous experiment, we indirectly demonstrate the effectiveness of \baby in modeling multi-domain behaviors through the Recall metric. In this section, we present two examples based on the \textbf{Diff}(introduced in Section\ref{sec:impact_data}) setting to visually show the effectiveness of textual information in multi-domain recommendation. These two are shown in Table ~\ref{tab:case_study}. In the first case, the historical interaction sequence mainly comes from the Office and the Scientific, with the user's next interaction domain being Arts. The behavior in the Office suggests that the user may need to purchase some supplies related to the office pen, hence the "eraser" items are included for recommendation by the large language model. Also, the behavior in the Scientific domain indicates that the user might be a doctor, so the next correct item - bandage shears, is successfully included in the top results. The second case is relatively more difficult to guess. The historical interaction sequence includes two items related to microphones, and the target item does not have a direct textual link to microphones. However, the target item is an audio processing product, which is usually used in conjunction with microphones. The ID-based method like SASRec can not well identify this semantic cue, since the relevant historical items are from different domains. Fortunately, the exploitation of large language models can address this kind of problem well.

\begin{figure}[htbp]
    \centering{
    \subfigure{\includegraphics[width=0.95\textwidth]{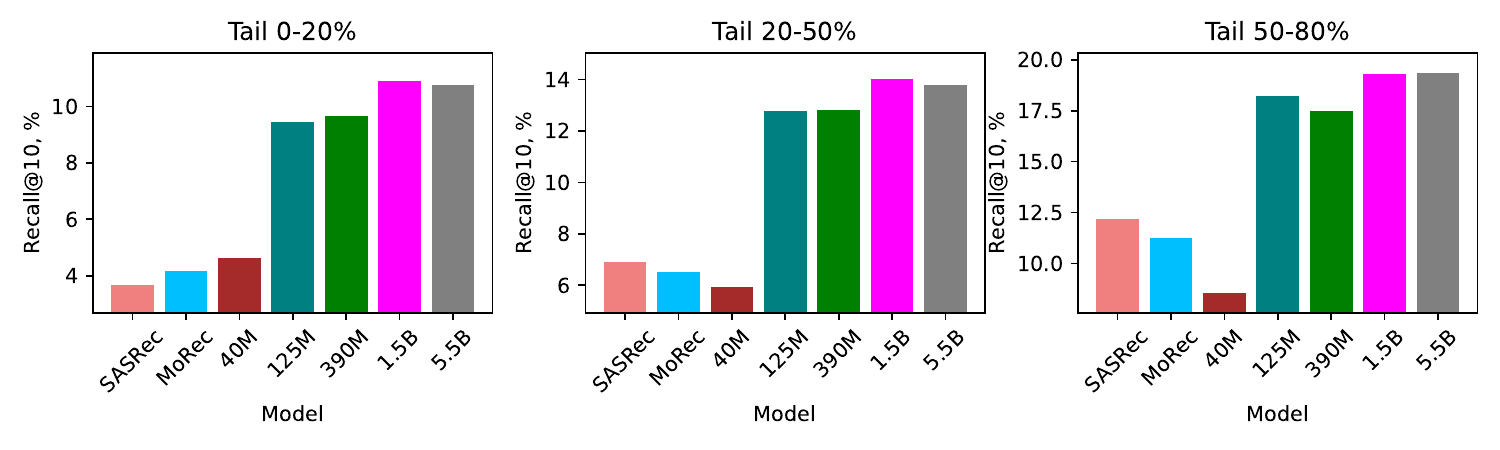}}
    \subfigure{\includegraphics[width=0.95\textwidth]{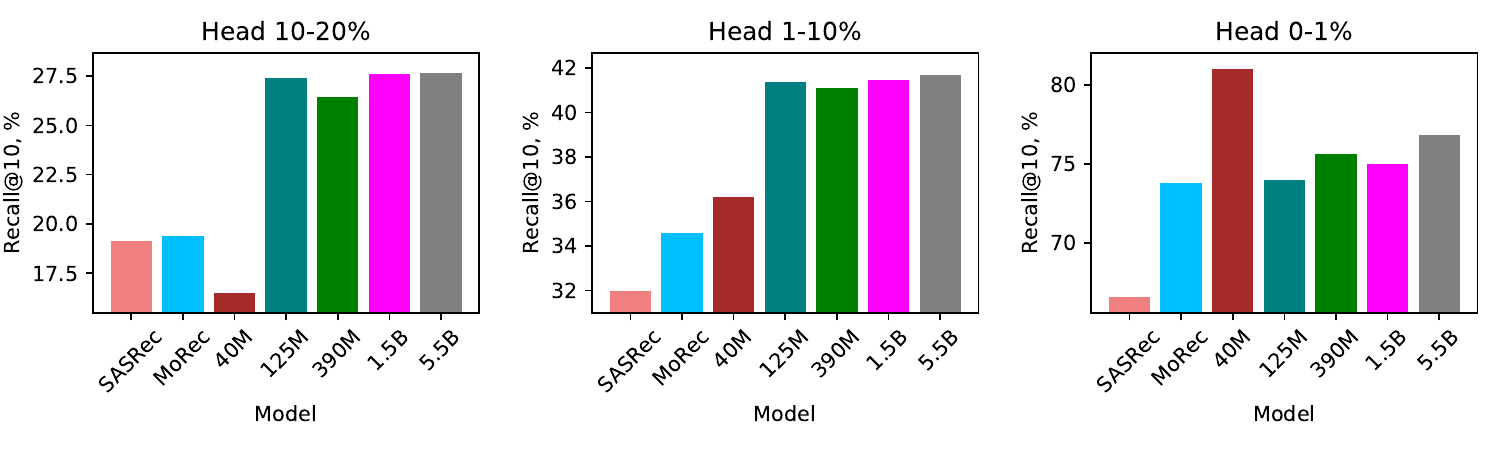}}
    }
    \caption{Performance of different frequency items. The backbone of \baby is FLAN-T5.}
    \label{fig:cold_start}
\end{figure}

\begin{figure}[htbp]
    \centering{
    \subfigure{\includegraphics[width=0.95\textwidth]{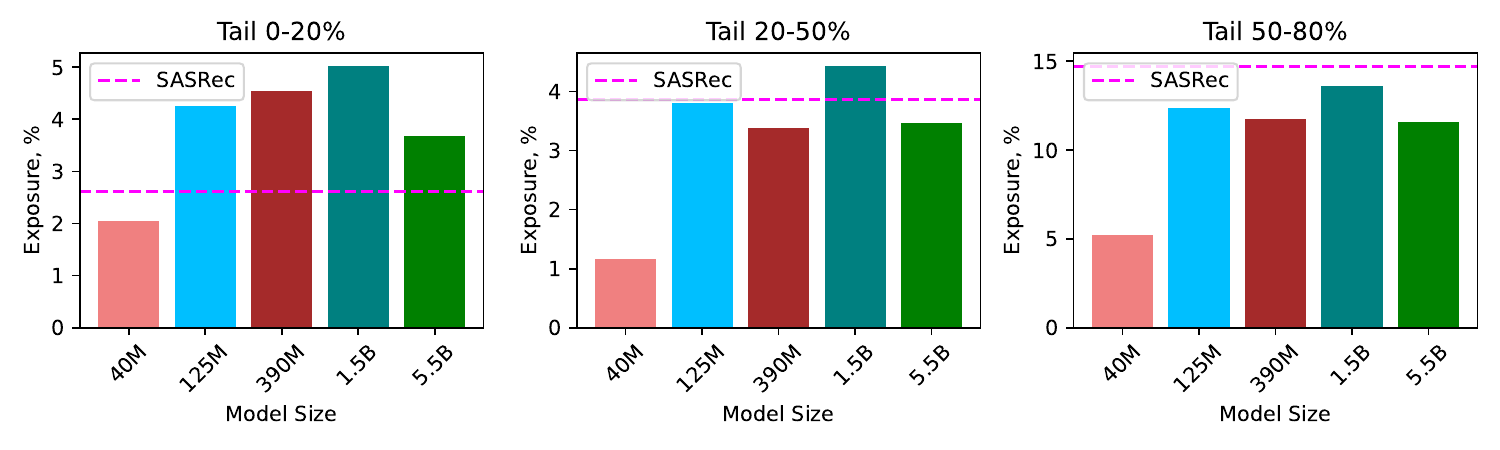}}
    \subfigure{\includegraphics[width=0.95\textwidth]{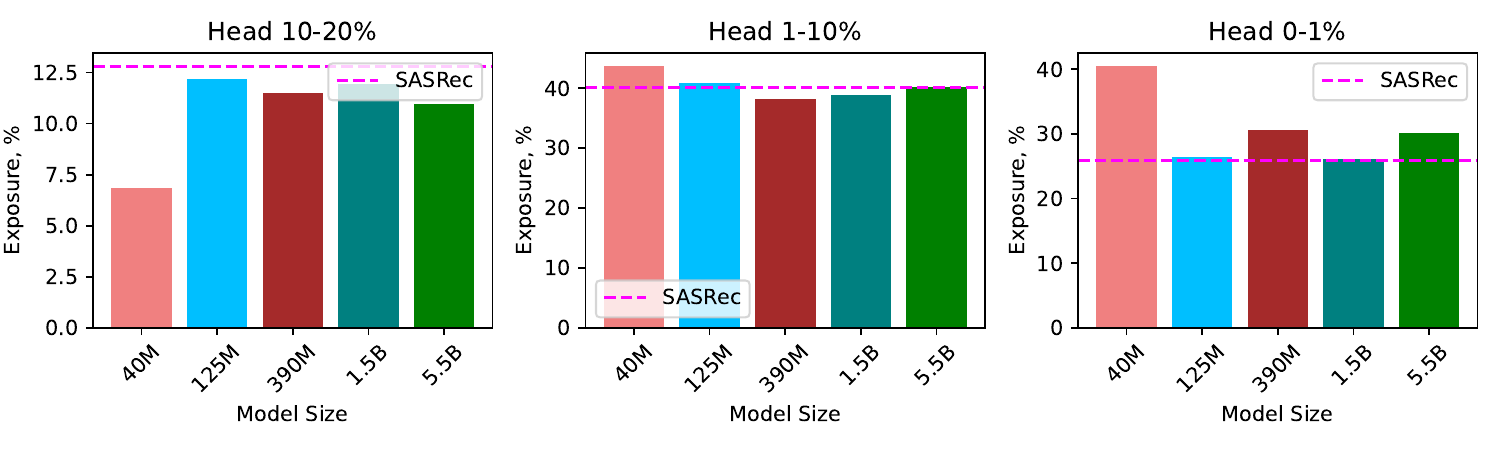}}
    }
    \caption{Exposure of different frequency items. The backbone of \baby is FLAN-T5.}
    \label{fig:item_fairness}
\end{figure}

\subsection{The Larger Language Model is better at capturing the semantic correlations. (Q2)}
\label{sec:Q2}
To answer this question, we conduct three types of experiments: cold-start item, Matthew Effect, and multi-domain item visualization. 
We utilize popularity bias as the proxy to analyze why \baby works: If a recommendation model relies more on item-level collaborative filtering information, then the model will be more severely affected by the Matthew effect and show a significant popularity bias. Conversely, if a recommendation system exhibits less popularity bias towards items and performs better on cold-start items, it will depend more on semantic understanding capabilities. This assumption is based on the notion that if recommendations are based on semantic understanding, items with similar semantics but different popularity would have comparable exposure, thus reducing popularity bias. 

\label{sec:cold_start}
\paragraph{Cold-start.} 
We first categorize all items based on their appearance frequency in the training set.
By sorting items in ascending order of frequency and taking the top 20\%, 50\%, and 80\% as the
tail items, and the bottom 1\%, 10\%, and 20\% as the head items, we calculate the performance for the different groups of items separately, in terms of Recall@10. The results are shown in Figure ~\ref{fig:cold_start}.
Firstly, for ID-based models, we observe that their effectiveness on long-tail items is notably inferior. For the bottom 20\% of items by frequency, the performance is smaller than 0.05. 
MoRec, which replaces ID representation with un-freezed text representation, doesn't achieve better long-tail performance. We believe the reason is that the utilization of text at the item-level makes it difficult to capture the relationships between texts.
In the case of the \baby, at a model size of 40M, the performance on low-frequency items is disappointingly low, even falling behind ID-based model, while the performance of the popular items is strangely high. This suggests that small-sized language models hold insufficient knowledge, leading to an especially overfitting for those high-frequency items. As the language model's size increases, the capability for handling low-frequency items progressively improves, while the performance for high-frequency items shows minimal variation, implying that popular items can be well modeled with a relatively small model size.  

\paragraph{Matthew Effect.}
It is commonly acknowledged that recommender systems are deeply influenced by the popularity bias, leading to the Matthew Effect phenomenon. These models tend to significantly overexpose high-frequency items at the expense of low-frequency ones. By employing language models to represent user sequences and item profiles, we aim to leverage richer semantic information to mitigate such biases. Therefore, we examine the exposure rates of items with varying degrees of popularity across both ID models and LLMRec.

\begin{figure}[thbp]
    \centering{
    \subfigure{\includegraphics[width=0.95\textwidth]{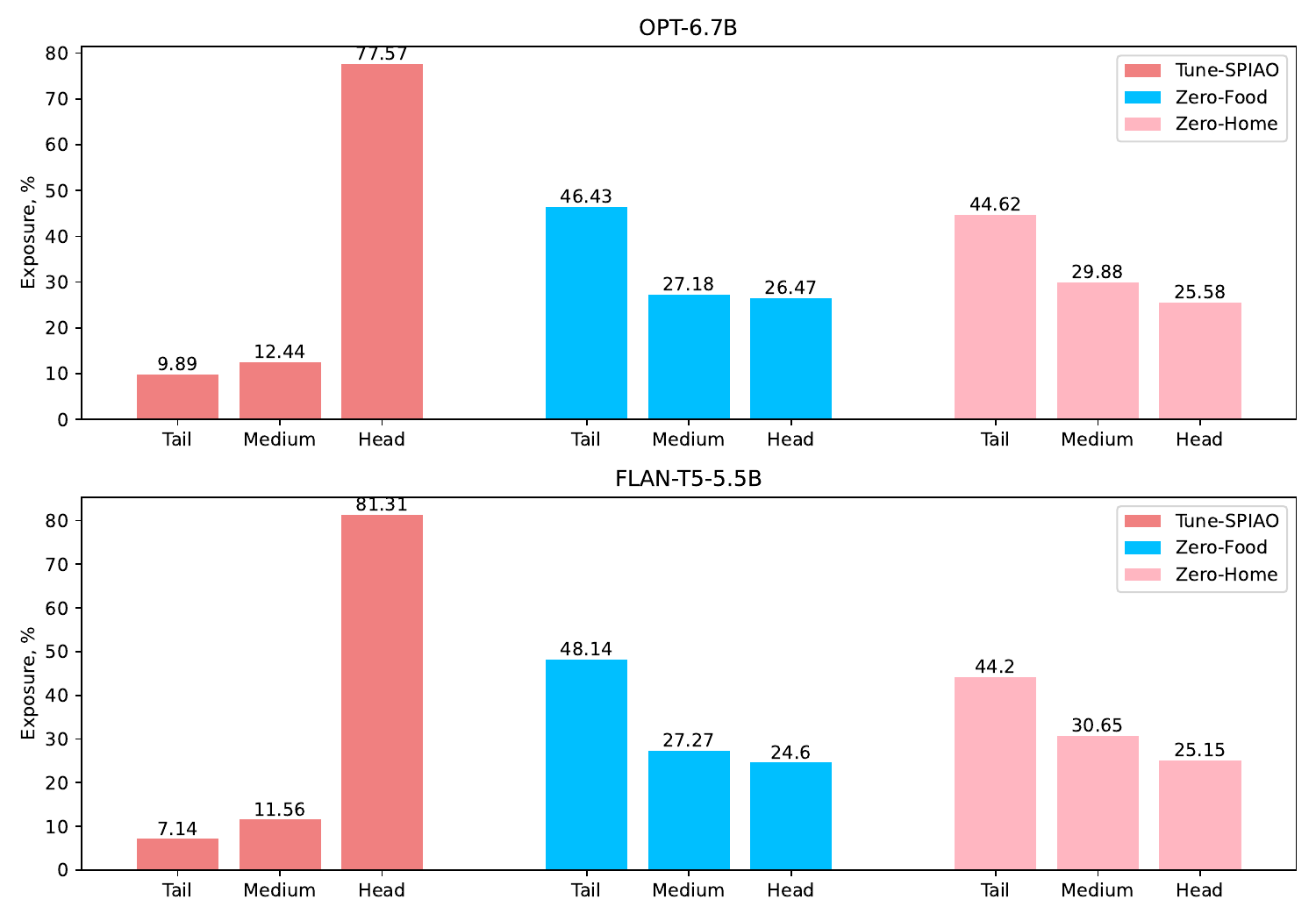}}
    }
    \caption{Exposure of items under different frequencies. Tail, Medium, and Head refer to the bottom 50\%, 50-80\%, and the most popular 20\% items in the training set respectively.}
    \label{fig:zero_fairness}
\end{figure}

The results are presented in Figure \ref{fig:item_fairness}. Surprisingly, All language models show a noticeable Matthew effect. Although they boost exposure for items of extremely low popularity (Tail 0-20\%) more than the ID model, the exposure rates for high and relatively lower-frequency items remain similar to those of the ID model. It demonstrates that language models mainly depend on item-level collaborative filtering, not token-level semantic correlations. To further validate our conclusion, we analyze whether \baby is affected by popularity bias in the zero-shot domain, where the model has received no training on the domain's data. Specifically, we select another two Amazon sub-categories: "Grocery and Gourmet Food" and "Home and Kitchen"~\footnote{Due to the computing source limitation, we sample $30\%$ sequences from the Food and Home datasets.}. The results, as shown in Figure ~\ref{fig:zero_fairness}, indicate that on the zero-shot domain, the exposure rate of items is almost unrelated to frequency, meaning that the language model is able to provide a relatively unbiased recommendation. Combining the results from Section ~\ref{subsubsec:zero_performance}, we can observe that language models are capable of capturing both item-level collaborative filtering correlations and semantic correlations between items through training, and the semantic correlations can be transferred to the zero-shot dataset with almost no popularity bias.

\paragraph{Multi-domain Visualization.}
\label{sec:visulization}
\begin{figure}[htbp]
    \centering{
    \subfigure{\includegraphics[width=0.32\textwidth]{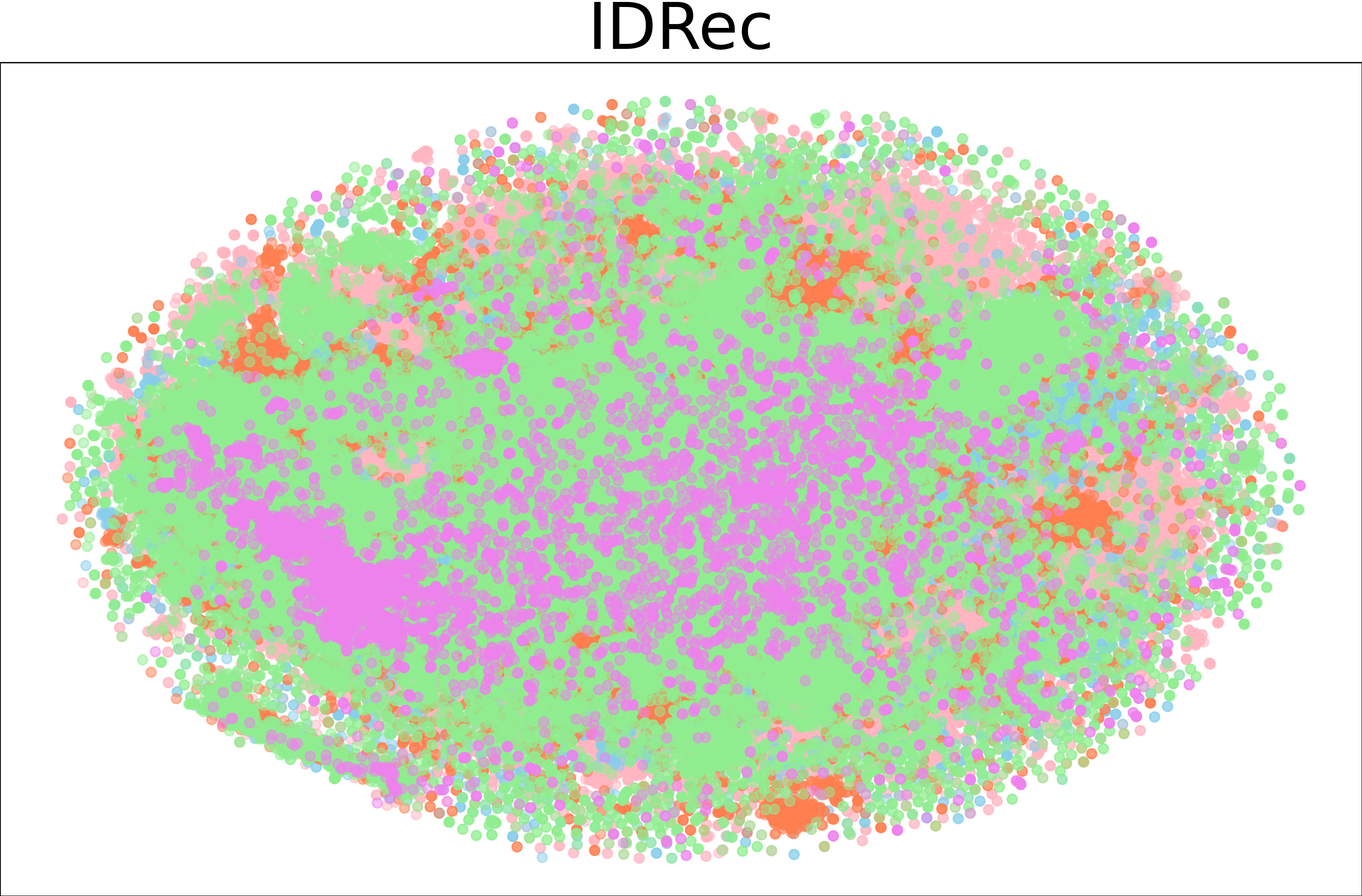}}
    \subfigure{\includegraphics[width=0.32\textwidth]{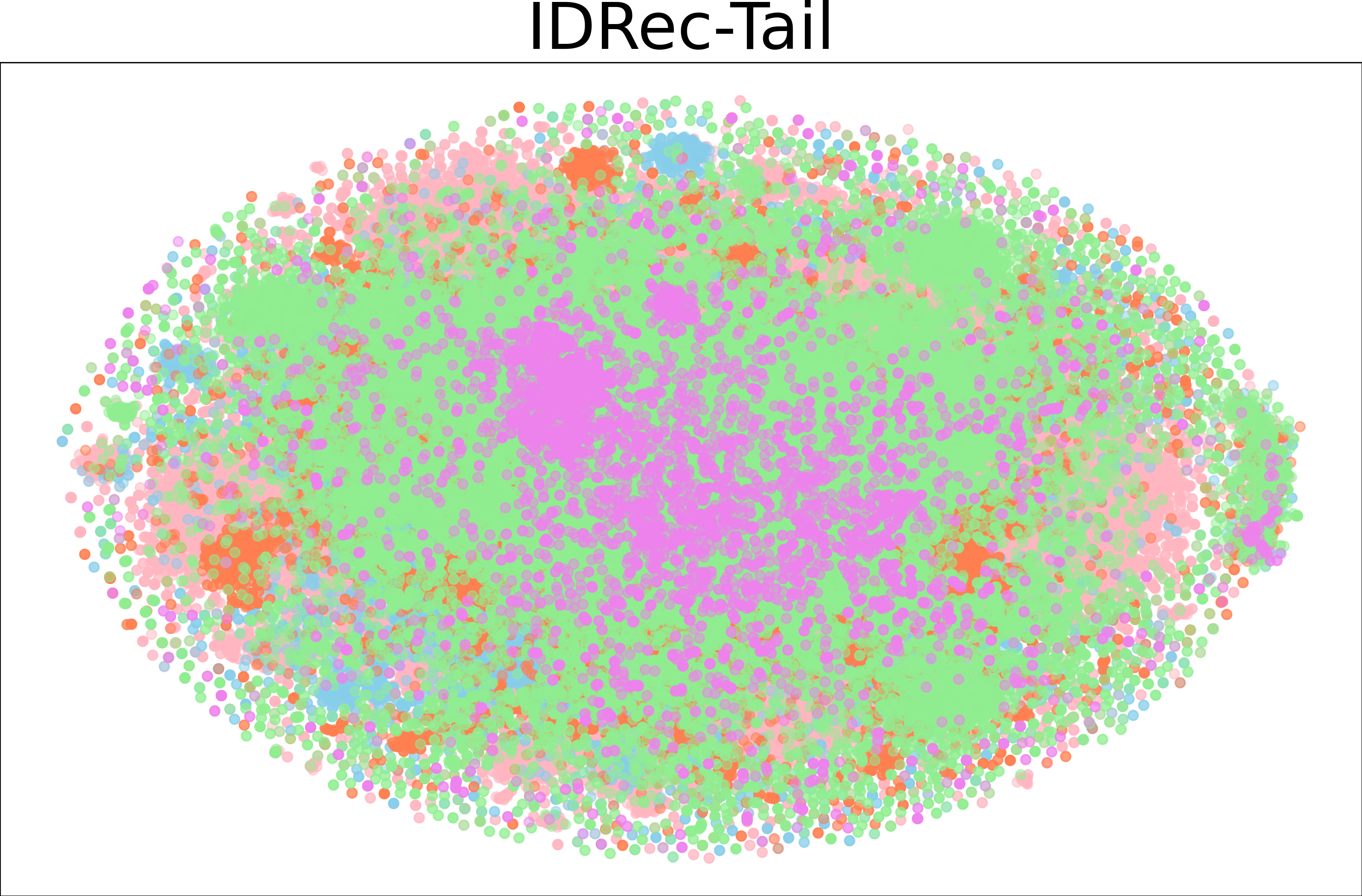}}
    \subfigure{\includegraphics[width=0.32\textwidth]{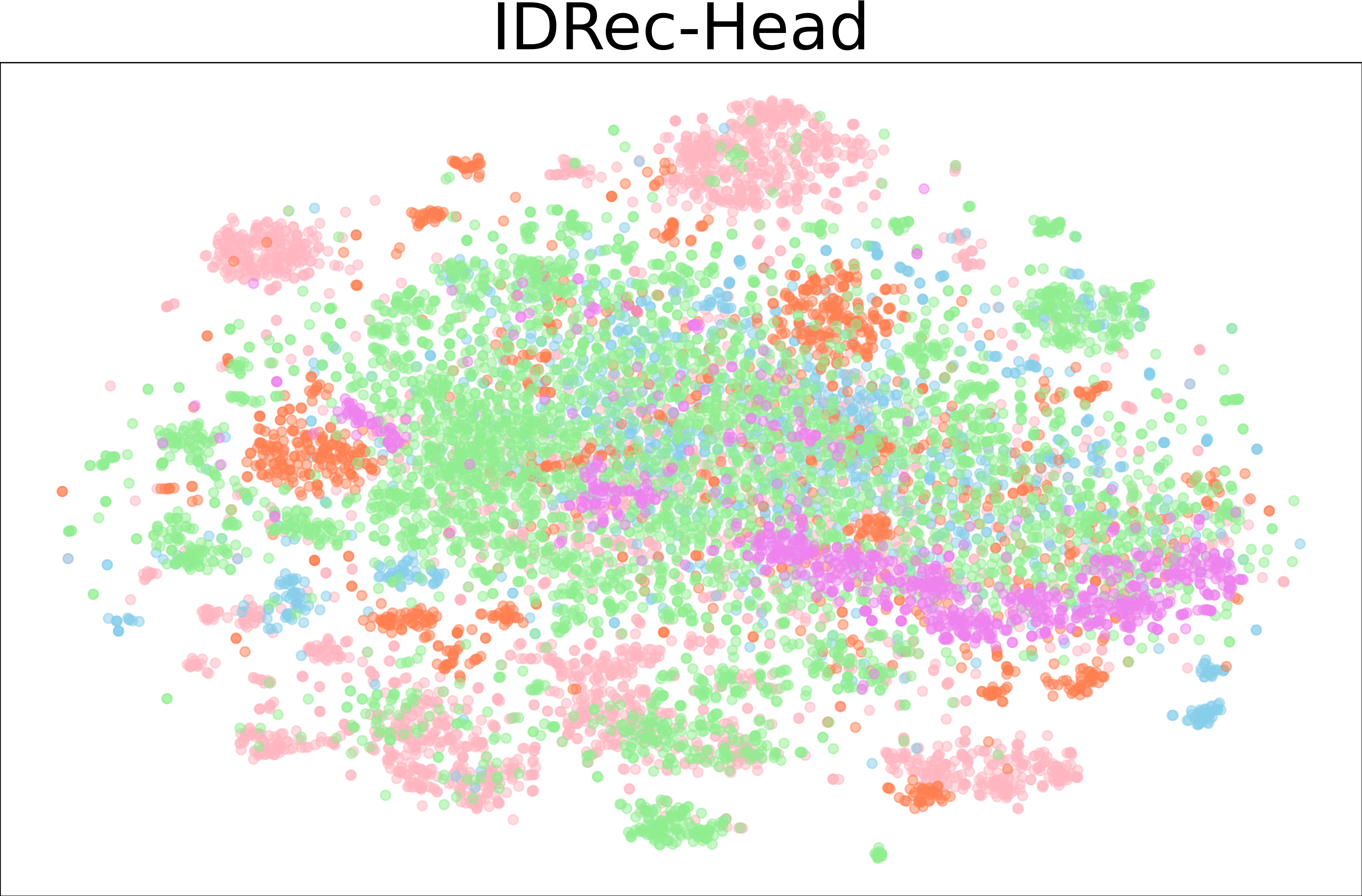}}
    \subfigure{\includegraphics[width=0.32\textwidth]{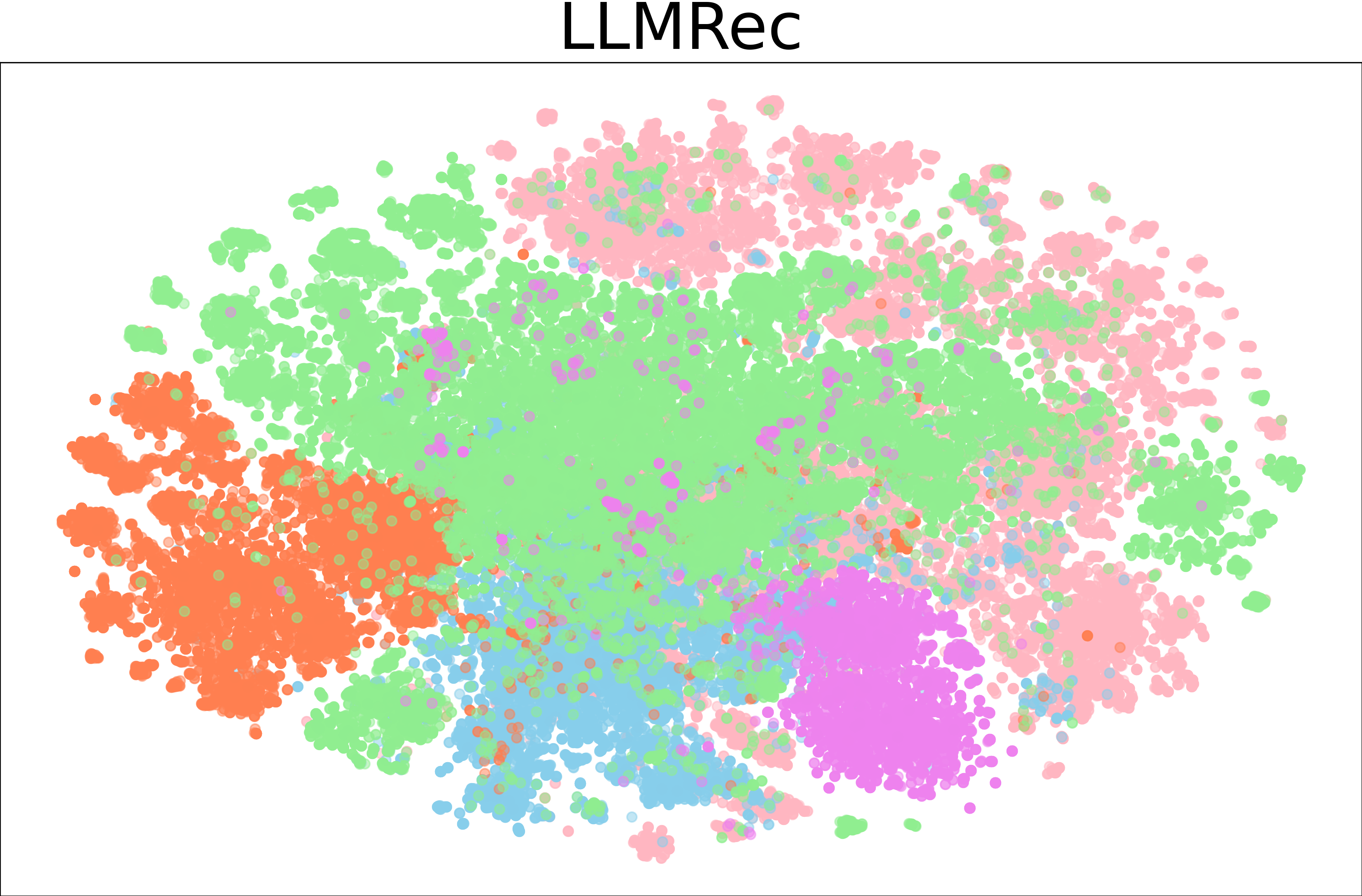}}
    \subfigure{\includegraphics[width=0.32\textwidth]{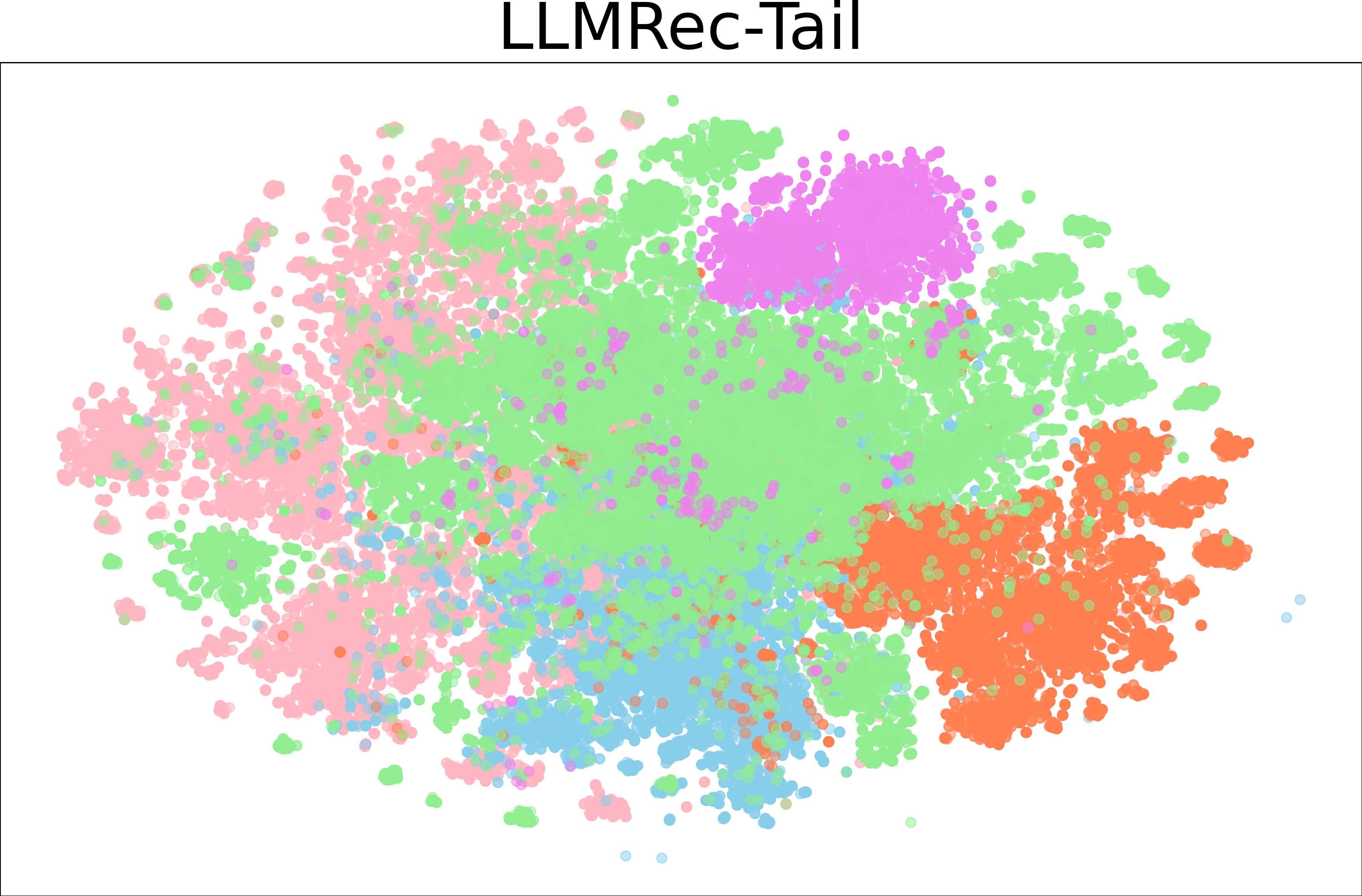}}
    \subfigure{\includegraphics[width=0.32\textwidth]{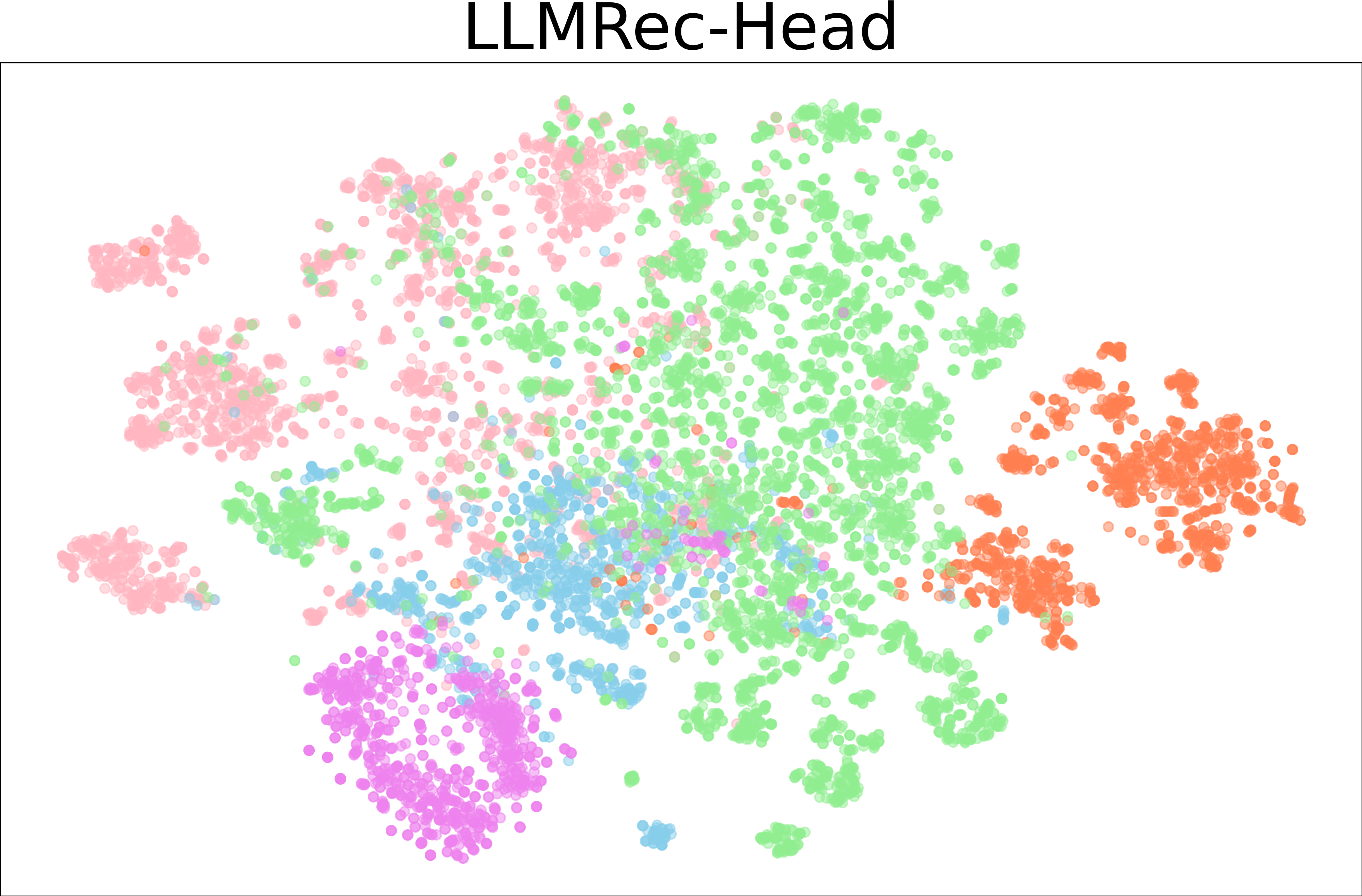}}
    \subfigure{\includegraphics[width=0.5\textwidth]{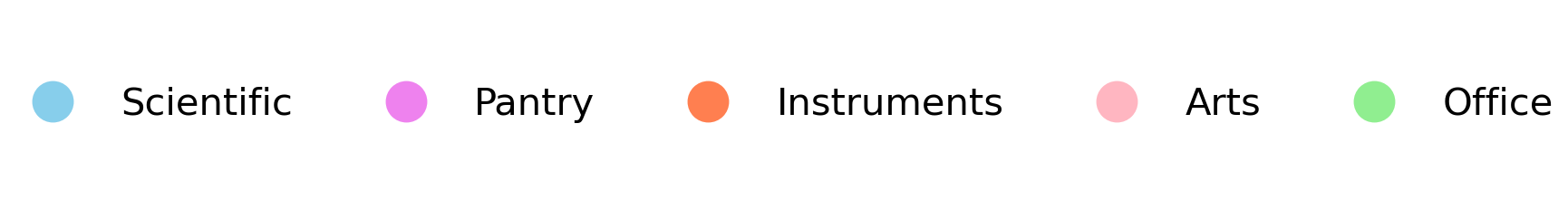}}
    }
    \caption{t-SNE visualization for IDRec(SASRec) model and \baby item representations, different colors represent different domains. "-Head" means that only Head items are retained for visualization, and "-Tail" means that only items are retained for visualization.}
    \label{fig:visulization}
\end{figure}
Because the limited interactions resulting in insufficient training for tail items, we believe it is difficult for the model to learn a good item representation, especially in a multi-domain recommendation scenario where behaviors across different domains are even sparser. Therefore, we conduct a visual analysis of the item representations learned by the ID-based models and the \baby models using t-SNE, as shown in Figure ~\ref{fig:visulization}. Furthermore, we only visualize the Head item and the Tail item to facilitate the analysis.
Here, we find that the representations learned purely by ID bear poor distinction across different domains, with all domains being highly mixed together. 
Upon further comparison between the Tail item and Head item representations, it is obvious that IDRec's Head items have better domain distinction, while Tail items are still intermixed. In contrast, the item representations learned by \baby possess a higher degree of distinction across different settings. These analyses demonstrate that \baby has a superior modeling capability for sparse data compared to IDRec, which is consistent with our motivation that the pre-trained large language models are effective in modeling users’ behaviors across different domains. The t-SNE visualization of only the top items in \baby showed even greater inter-class distance. It is worth noting that we only used the item title text and item category text is not used.

Combining the observations above, we find the reasons behind \baby's better performance are somewhat complex. Its popularity bias on extremely popular items demonstrates that \baby still heavily relies on item-level collaborative filtering information for recommendations. However, the significant increase in exposure and performance improvement on long-tail items, along with across-domain performance enhancement, suggests that it also leverages the semantic understanding capabilities of language models. Furthermore, the more pronounced performance improvement on tail items by larger models indicates the larger language model is better at capturing the semantic correlations.

\subsection{Advanced Language Model Techniques for Recommendation(Q3)}

\subsubsection{The Impact of Language Model Size}
\label{sec:model_size}
\begin{figure}[htbp]
    \centering{
    \subfigure{\includegraphics[width=0.90\textwidth]{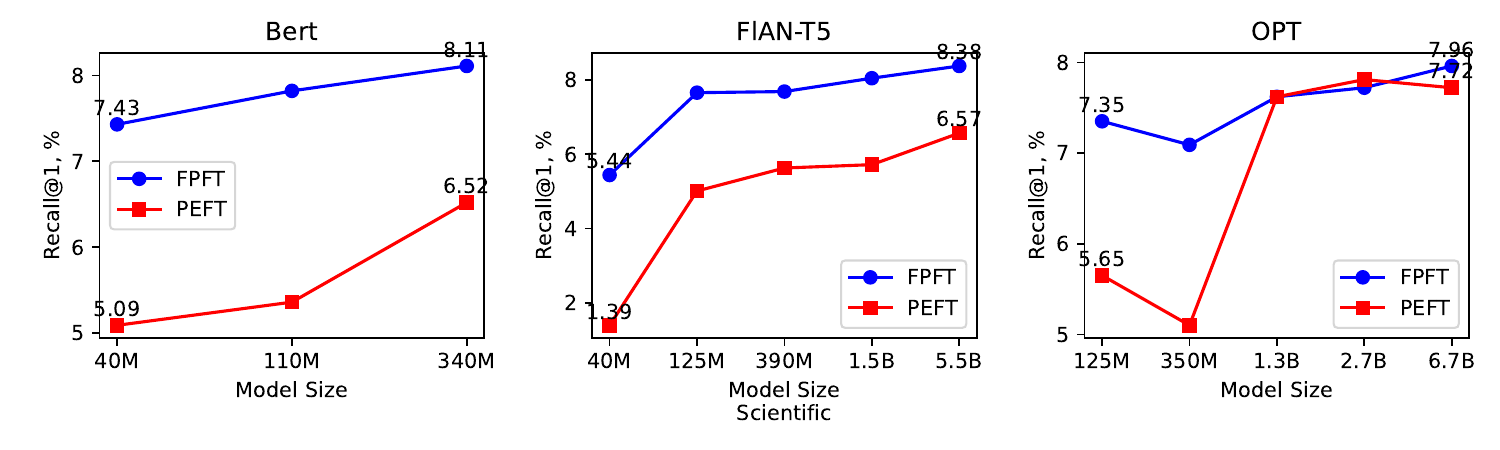}}
    \subfigure{\includegraphics[width=0.90\textwidth]{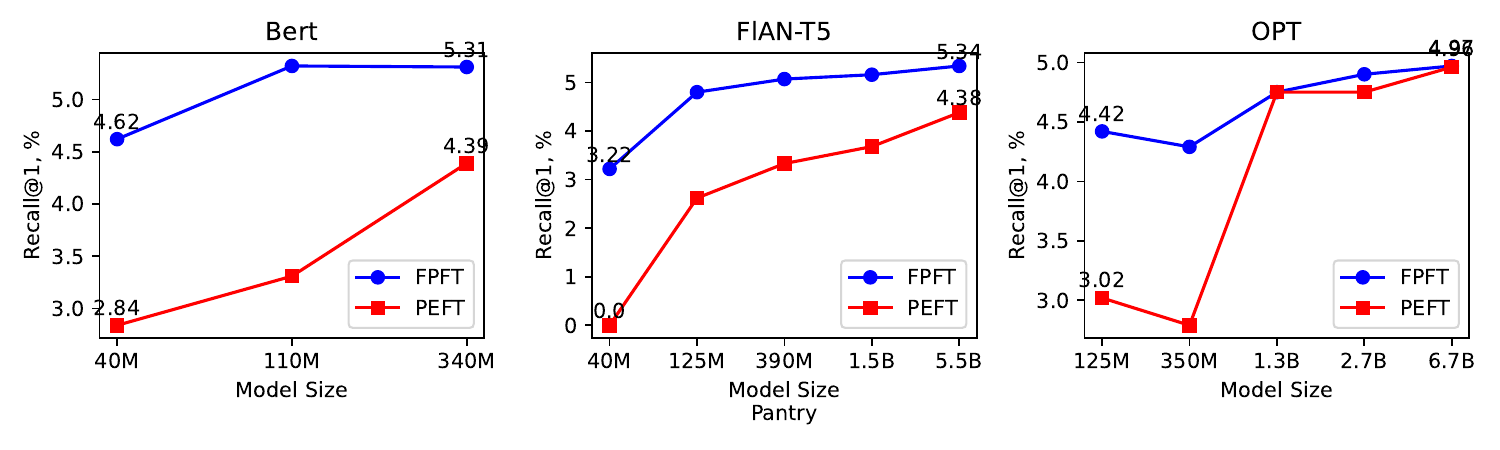}}
    \subfigure{\includegraphics[width=0.90\textwidth]
    {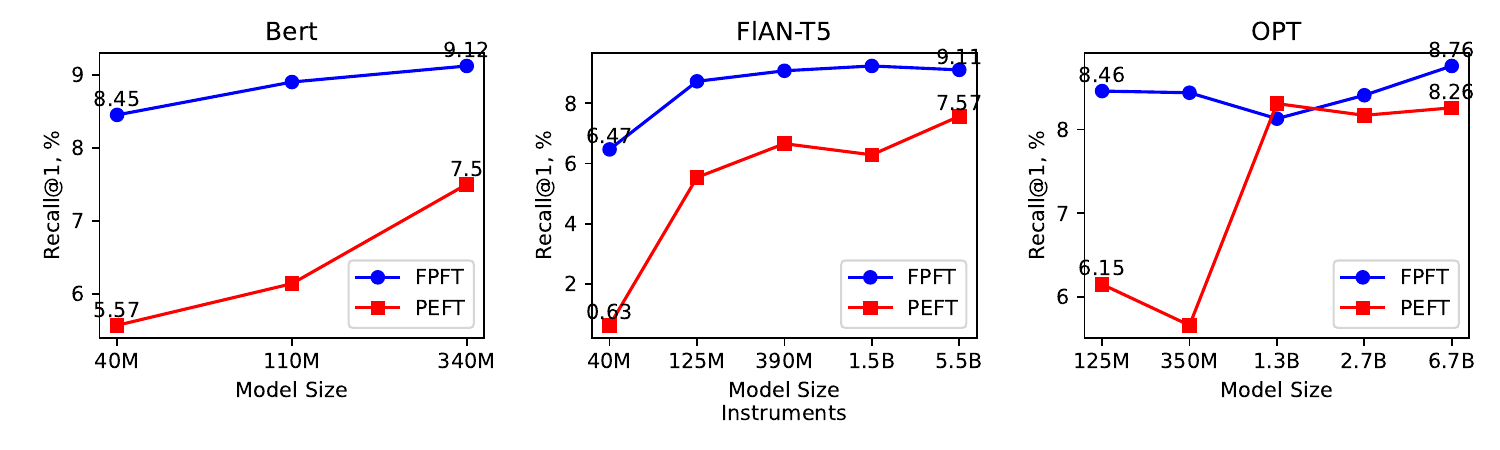}}
    \subfigure{\includegraphics[width=0.90\textwidth]{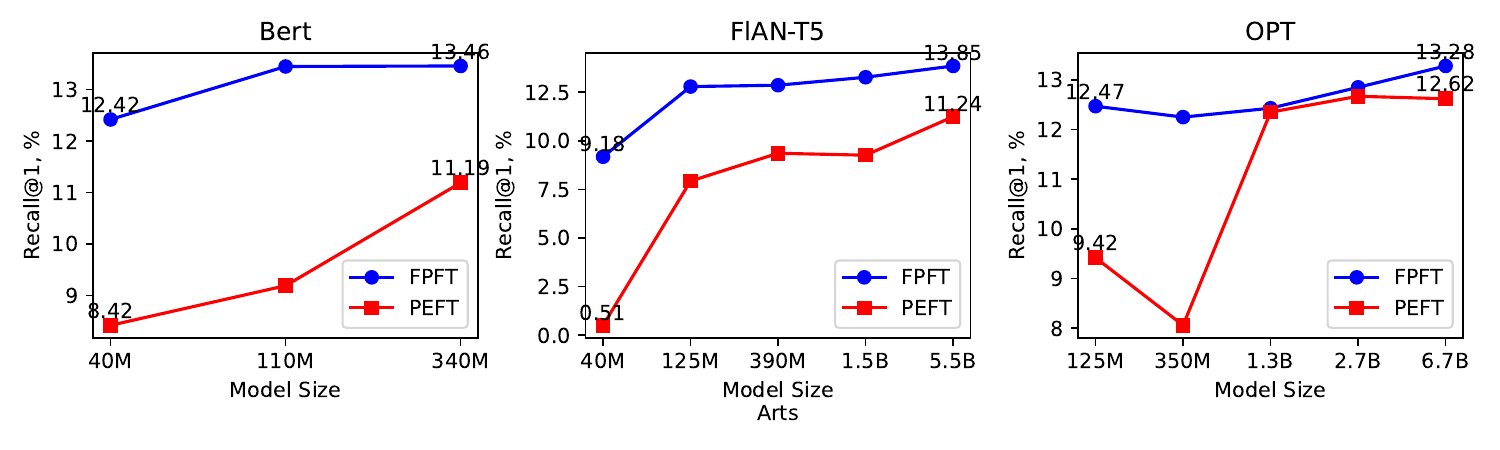}}
    \subfigure{\includegraphics[width=0.90\textwidth]{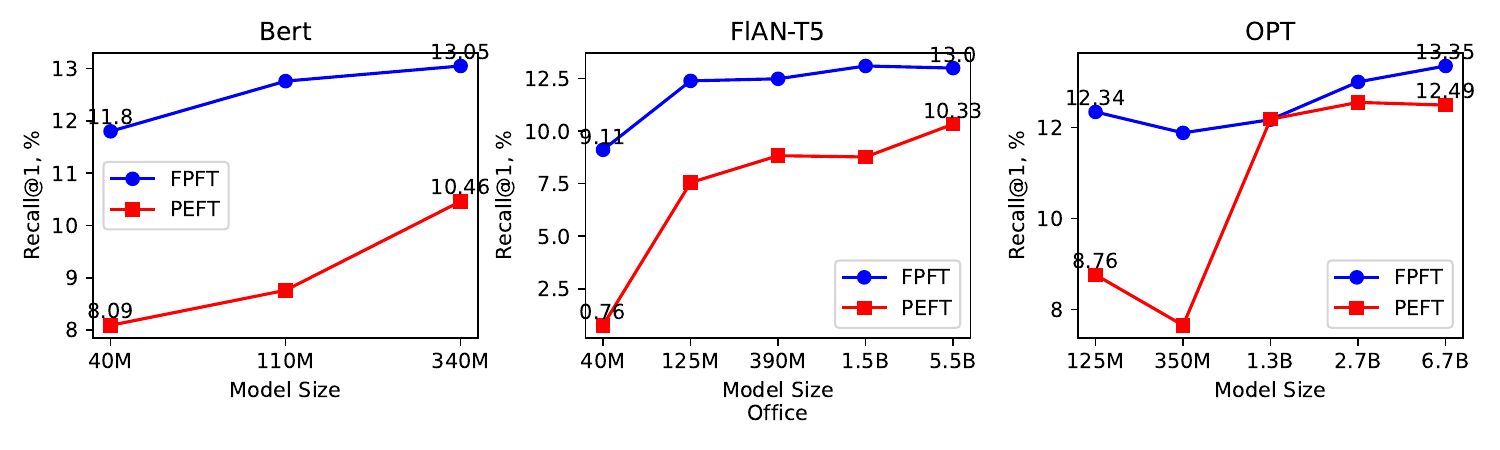}}
    }
    \caption{Results of various sizes, backbones and fine-tuning methods. The FPFT and PEFT represent full parameter fine-tuning and parameter efficient fine-tuning respectively.}
    \label{fig:model_size_tuning_methods}
\end{figure}

Notes that a larger pre-trained base model is more competent for more complicated downstream tasks and acquires significant improvement in NLP and CV fields. In Section~\ref{sec:perf}, we have already manifested that $110M$ around parameters could achieve good performance in most scenarios. Thus, in this section, we further investigate the impact of model size on the final performance. To this end, we dedicate various model sizes for different backbones, i.e., ranging from $40M$ BERT-Medium to $6.7$B OPT model. Figure ~\ref{fig:model_size_tuning_methods} presents the results across different sizes.
 
\begin{figure}[t]
    \centering{
    \subfigure{\includegraphics[width=0.45\textwidth]{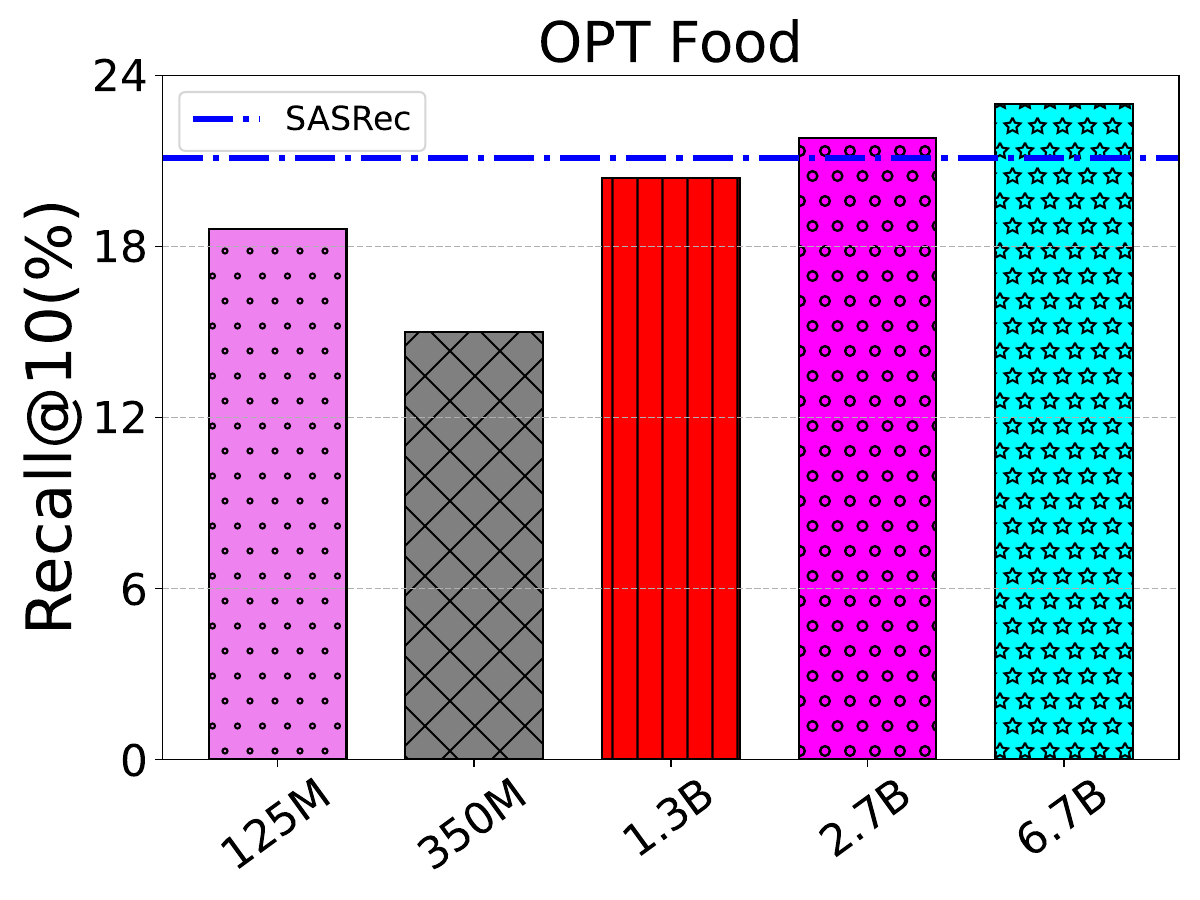}}
    \subfigure{\includegraphics[width=0.45\textwidth]{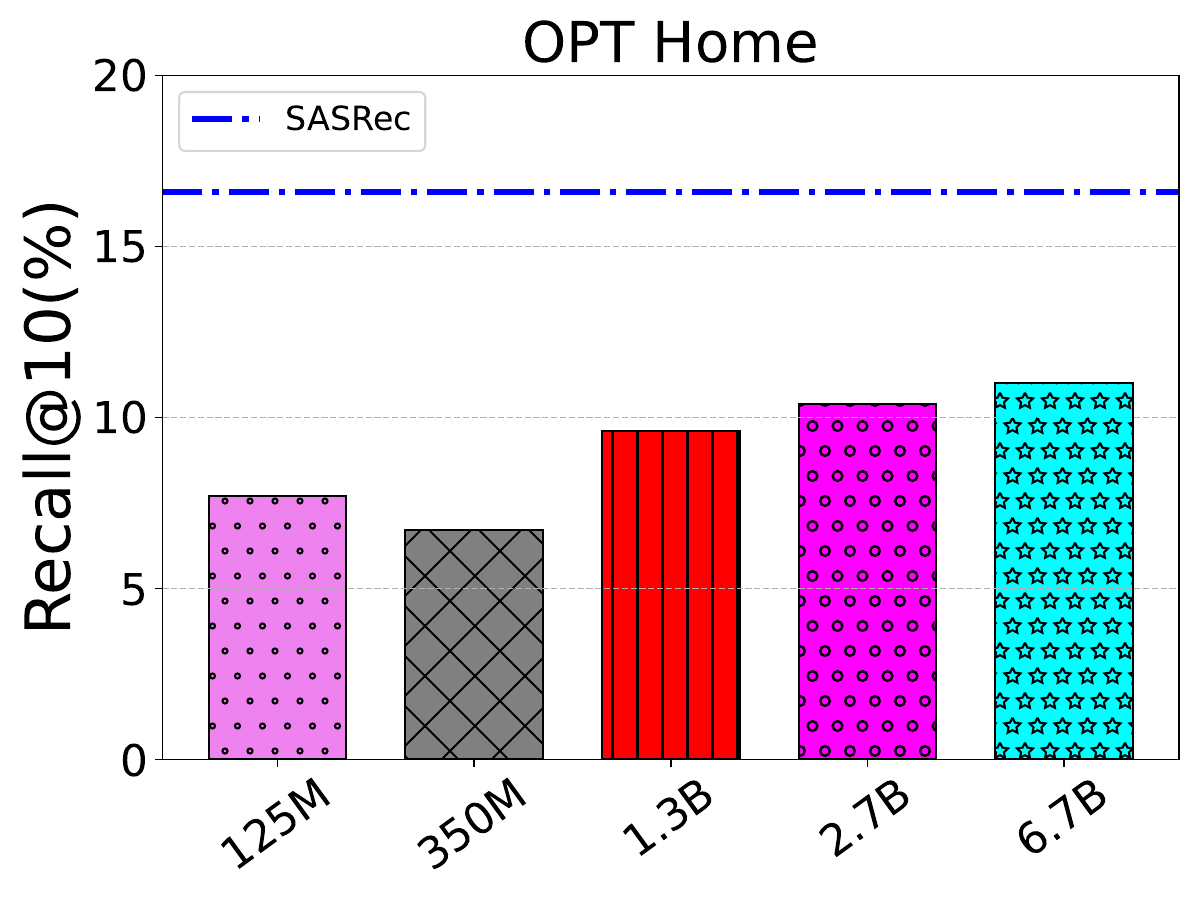}}
    \quad
    \subfigure{\includegraphics[width=0.45\textwidth]{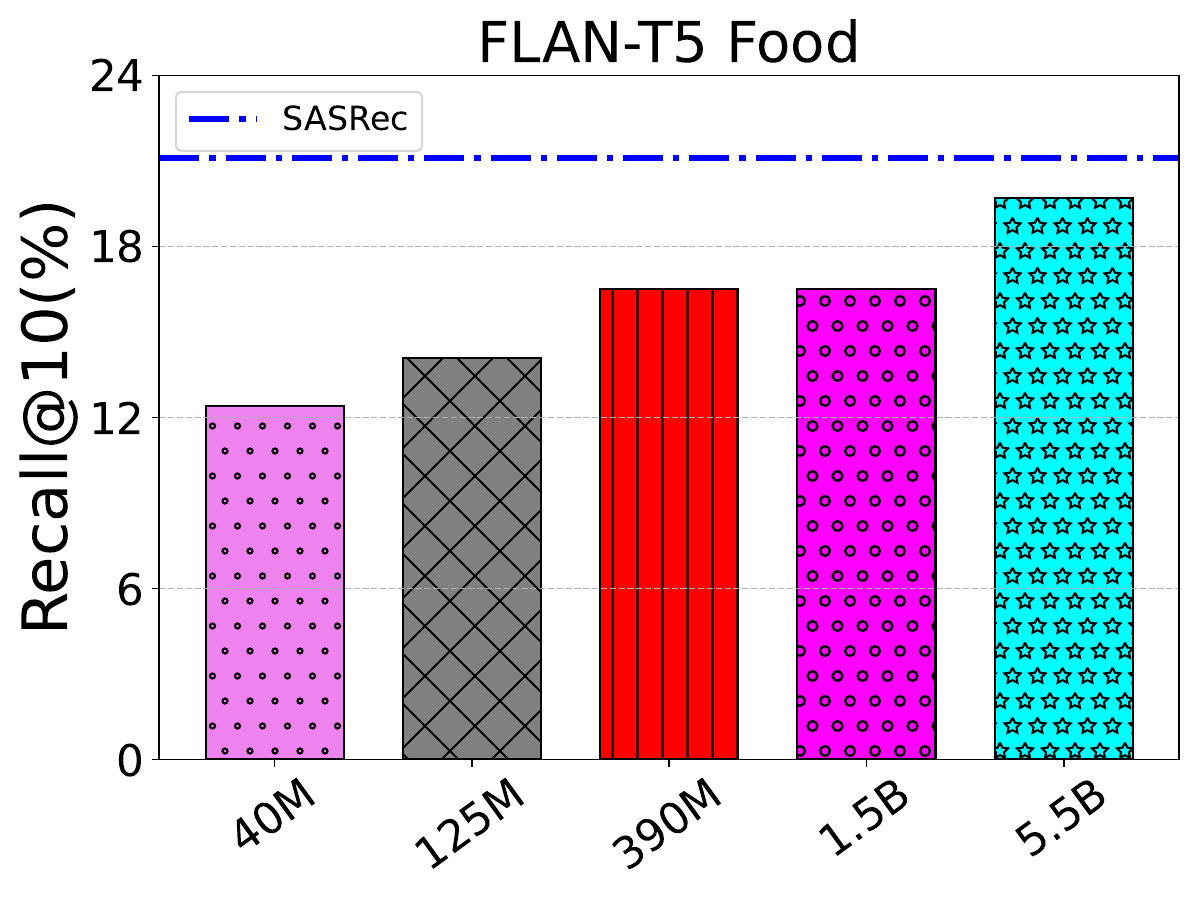}}
    \subfigure{\includegraphics[width=0.45\textwidth]{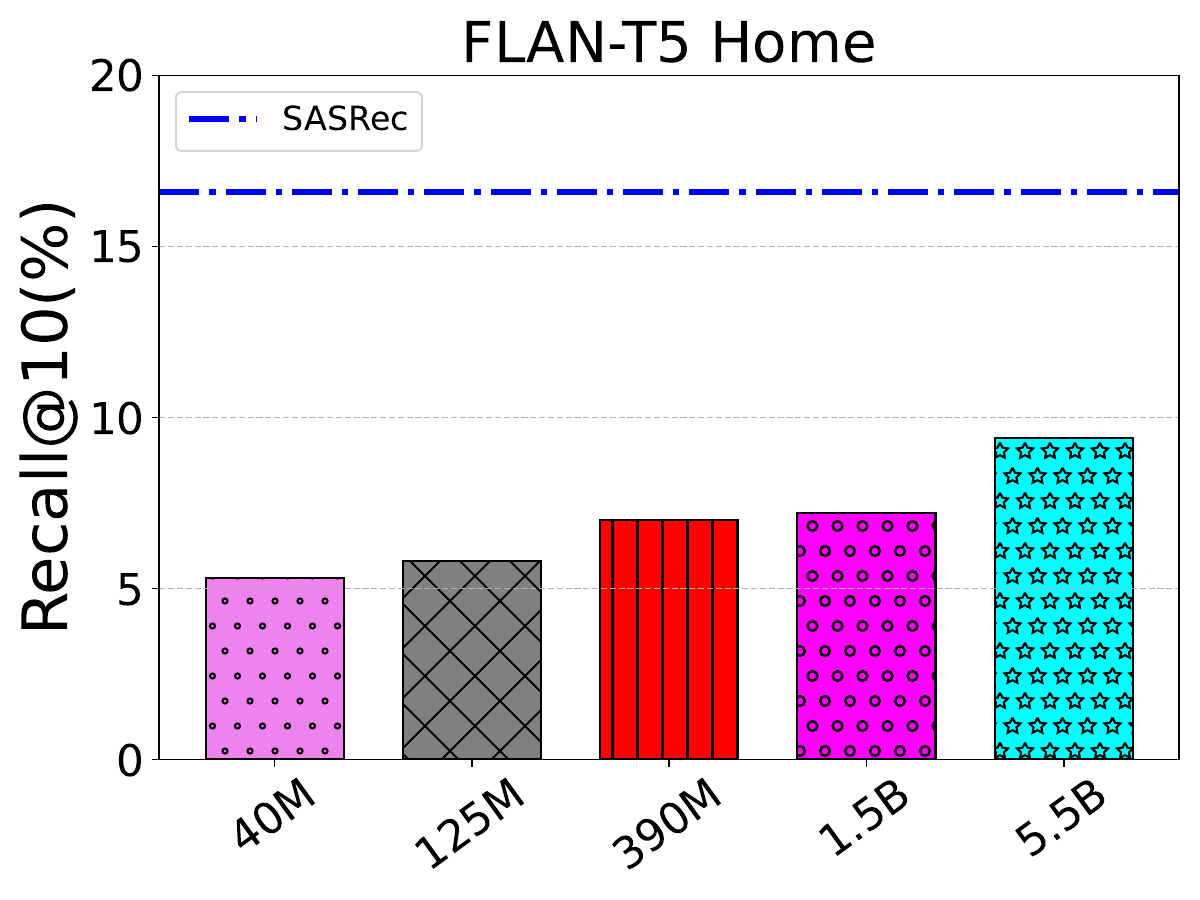}}
    }
    \caption{Zero-shot domain performance. }
    \label{fig:zero}
\end{figure}

\paragraph{Fine-tune Domain Performance}
Observing Figure~\ref{fig:model_size_tuning_methods} we could find that by increasing parameters the performance will acquire a huge gain when the model size is rather limited, i.e., $10\%$ and $33\%$ relative increase in Recall@1 when the size is expanded from $40M$ to $110M$ for BERT and $40M$ to $125M$ for FLAN-T5, respectively. However, this trend will not remain when the model size exceeds $100M$. For instance, in terms of OPT, the minimum size ($125M$) presumably is sufficient to achieve near-optimal performance, and when we increase the model size to $6.7B$, the performance improvement is relatively small.
A similar phenomenon could also be observed in the BERT and FLAN-T5 models, i.e., BERT-$330M$ does not show sufficient superiority to the BERT-$110M$, and the same holds true for FLAN-T5.(ten times increasing of model size only achieves $2\%$ to $5\%$ relative improvements). Consequently, we believe it will be difficult to acquire a huge performance improvement when the model exceeds a certain size ($100M$ around), which is quite different from the observations made in NLP and CV tasks. 

\paragraph{Zero-shot Domain Performance}
\label{subsubsec:zero_performance}
Considering that the model size can also facilitate the generalization ability of LLM. We further analyze the zero-shot performance with different model sizes on unseen domains: Food and Home. Experimental results are reported in Figure ~\ref{fig:zero}. We could find that as the model size increases, the zero-shot performance can be greatly improved, except for OPT-350M~\footnote{Researchers in the NLP community have also found that the performance of OPT-350M is poor. The reason given by HuggingFace members is that there maybe a bug with the 350M weights, the related URL is https://github.com/huggingface/transformers/issues/17653.}. Encouragingly, on Food dataset, OPT-6.7B even outperforms SASRec, a strong baseline that is well-trained via item ID. Moreover, we find that the model structures will have an overwhelming impact on zero-shot performance against model size, i.e., the performance of OPT is much better than FLAN-T5 when their model size is under the same order of magnitude.

Based on the above results, we can see that "bigger is better" can not hold well in the multi-domain recommendation scenarios. Combining the observations in Section ~\ref{sec:Q2}, we believe the possible reasons would be as follows: 1) Zero-shot experiments indicate that larger models possess more universal knowledge, with the zero-shot performance significantly superior to that of smaller ones. 2) Different from NLP and CV tasks, the complicated patterns encapsulated in the collaborative filtering signals might be the recipe for success in recommendation systems. 3) However, model size seem to play a minor role in collaborative filtering signal modeling when it exceeds a certain size; a bigger model will not make learning the collaborative filtering signal easier.

\subsubsection{The Impact of Tuning Method}
\label{sec:tuning_method}
In Section~\ref{sec:model_size}, we show the results of full parameter fine-tuning (FPFT) for the pre-trained language models. However, as the model size increases, the computational cost for FPFT increases dramatically. For instance, fine-tuning the OPT-1.3B pre-trained model with AMP (Automatic Mixed Precision) requires at least $20$GB GPU memory. To circumvent this issue, parameter efficient fine-tuning (PEFT), i.e., fine-tuning very small groups of parameters to approach the performance of full parameter tuning, has become a prominent solution and also a cutting-edge research problem for LLM fine-tuning very recently. We, thereby, apply a widely-used PEFT method, i.e., LoRA~\cite{Lora} with low-rank $R=32$, for model fine-tuning and analyze the impact of tuning methods on the final performance. 
\paragraph{Parameter Efficient Fine-tuning Results.}
According to the results plotted in Figure ~\ref{fig:model_size_tuning_methods}, we can find that compared with FPFT, the performance of LoRA dropped substantially when the parameter is less than one billion. However, as the model size increases, the gap between them decreases progressively, indicating that it is essential to use FPFT for better recommendation performance when the model size is relatively small.
For models larger than 1B, PEFT is a viable option for more efficient model optimization for OPT series models, but the FLAN-T5 series models still have a significant decline.
Overall, for small models(<1B), PEFT is not suitable for the recommendation domain; for large models, the effectiveness of PEFT is related to the model architecture. This finding does not align with the conclusions drawn from applying PEFT to NLP tasks. In NLP tasks, even using PEFT in small language models (such as $T5_{base}$) can still yield comparable performance to FPFT~\cite{peft_survey}.

\begin{figure}[t]
    \centering
    \subfigure{\includegraphics[width=0.45\textwidth]{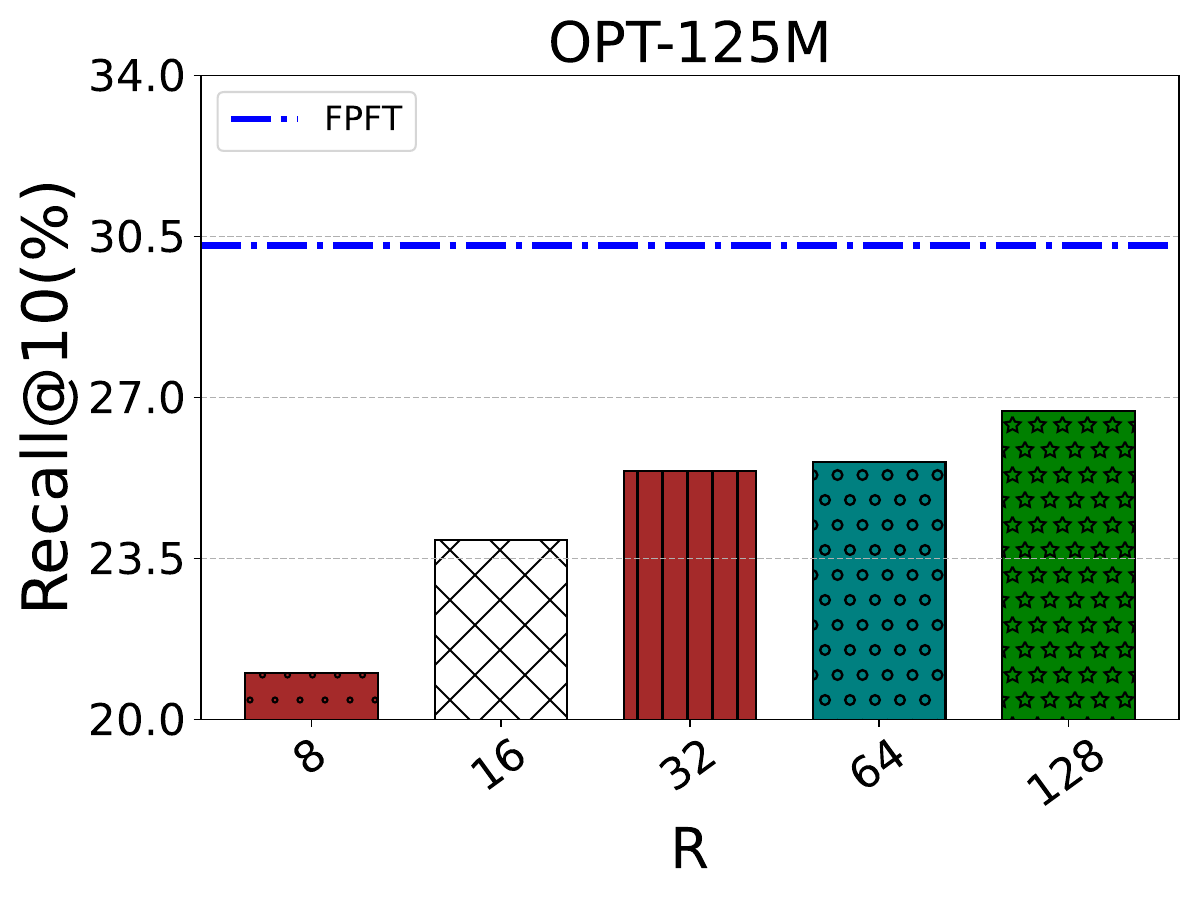}}
    \subfigure{\includegraphics[width=0.45\textwidth]{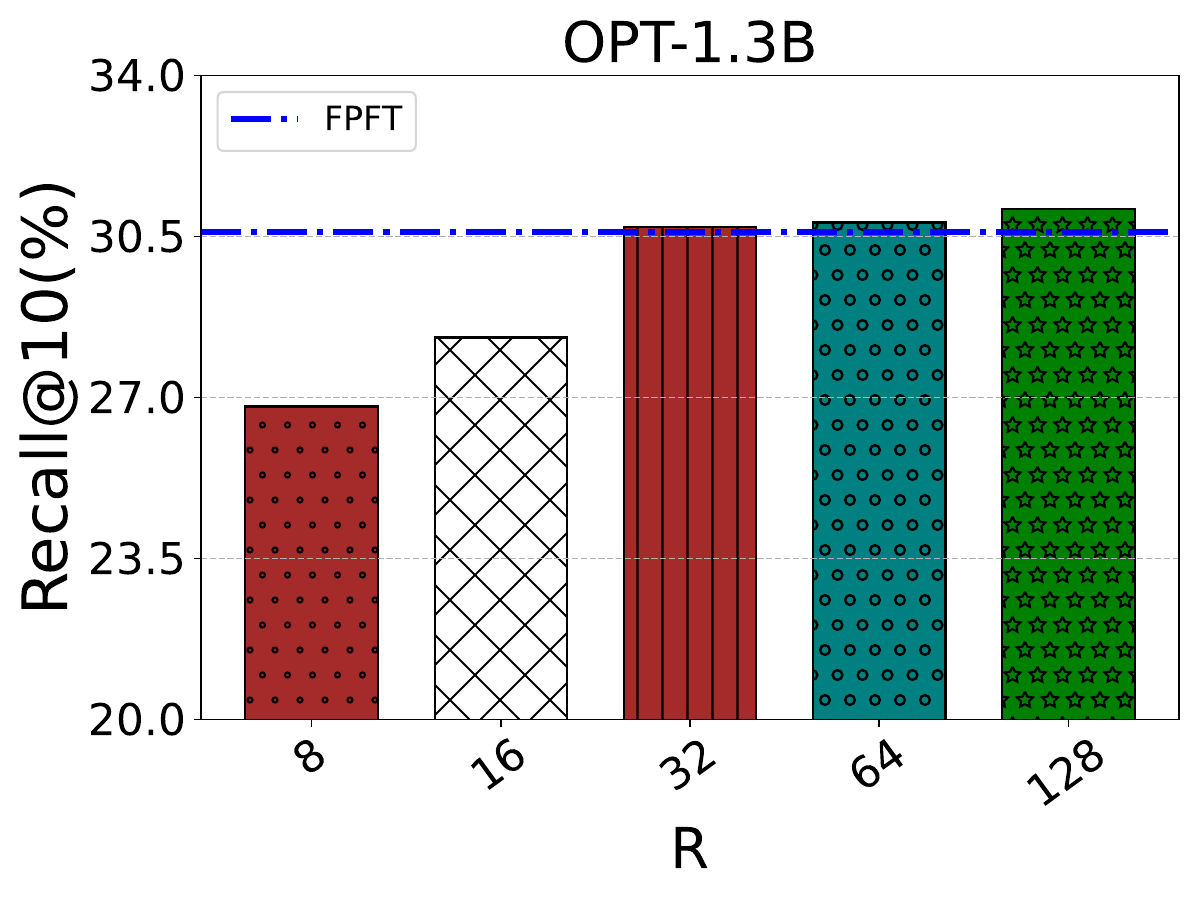}}
    \caption{The impact of trainable parameters in parameter efficient fine tuning}
    \label{fig:lora_r}
\end{figure}

\paragraph{The Impact of Trainable Parameters.}
To further explore the impact of the number of trainable parameters on the performance in PEFT, we validate the recommendation performance of different $R$ values in LoRA, the experimental results are shown in Figure ~\ref{fig:lora_r}. 
From the results, we can draw the following conclusions: 1) Overall, the model will present a better performance as trainable parameters increase. 2) Besides, although we increase the number of trainable parameters, PEFT is still inferior to the FPFT until the model size rises to one billion.
We believe the reason might be that the recommendation tasks (i.e., next-item prediction) and relevant corpus are deviated from NLP scenarios (e.g., text generation). Meanwhile, the generalization ability of small models is relatively limited, thus, the benefit of fine-tuning a small number of parameters for small model will be little.

\begin{table}[ht]
    \centering
    \caption{(\textbf{Left})Training cost of large different models on 6*A100-80G.  "h" represents the hour, "DDP" represents distributed data-parallel, "$Zero_{2}$" means deepspeed zero stage 2 parallel method and OOM represents out of memory in single A100-80G. The average training sequence token length of OPT is 122. SASRec is trained on a single A100-80G. \textbf{(Right)} Inference cost of different sequential models, we only considered the delay during model inference. We use OPT as the backbone of \baby}
    \begin{minipage}[b]{0.45\textwidth}
        \centering
        \begin{tabular}{c|ccc|c}
        \bottomrule
        Model                                     & Method  & Time  & Strategy      & Min-VMEM   \\ \hline
        SASRec                                    & -       & 1.3h    & -             & -  \\  \hline
        OPT-125M                                  & FPFT    & 8h    & DDP           & 16G  \\  \hline
        \multirow{2}{*}{OPT-1.3B    }             & FPFT    & 48h   & DDP           & 28G  \\ 
                                                  & PEFT    & 109h  & DDP           & 13G\\ \hline

        \multirow{2}{*}{OPT-2.7B    }             & FPFT    & 61h   & $Zero_{2}$    & 55G \\ 
                                                  & PEFT    & 81h   & DDP           & 24G \\ \hline
                                                            
        \multirow{2}{*}{OPT-6.7B    }             & FPFT    & 79h   & $Zero_{2}$    & OOM\\ 
                                                  & PEFT    & 152h  & DDP           & 56G\\ 
        
        \bottomrule
        \end{tabular}
    \end{minipage}
    \hfill
    \begin{minipage}[b]{0.40\textwidth}
        \centering
        \begin{tabular}{c|cc}
        \bottomrule
        Batch Size                                  & Model         & Time    \\ \hline
        \multirow{4}{*}{1    }                      & GRU4Rec       & 0.35ms  \\ 
                                                    & SASRec        & 1.1ms  \\ 
                                                    & OPT-125M      & 7.3ms    \\ 
                                                    & OPT-1.3B      & 17.3ms  \\ \hline

        \multirow{4}{*}{Maximum}                    & GRU4Rec       & 0.03ms    \\ 
                                                    & SASRec        & 0.045ms  \\ 
                                                    & OPT-125M      & 0.94ms    \\ 
                                                    & OPT-1.3B      & 7.0ms  \\

        \bottomrule
        \end{tabular}
    \end{minipage}
    \label{tab:training_cost}
\end{table}

\begin{figure}[htbp]
    \centering{
    \subfigure{\includegraphics[width=0.4\textwidth]{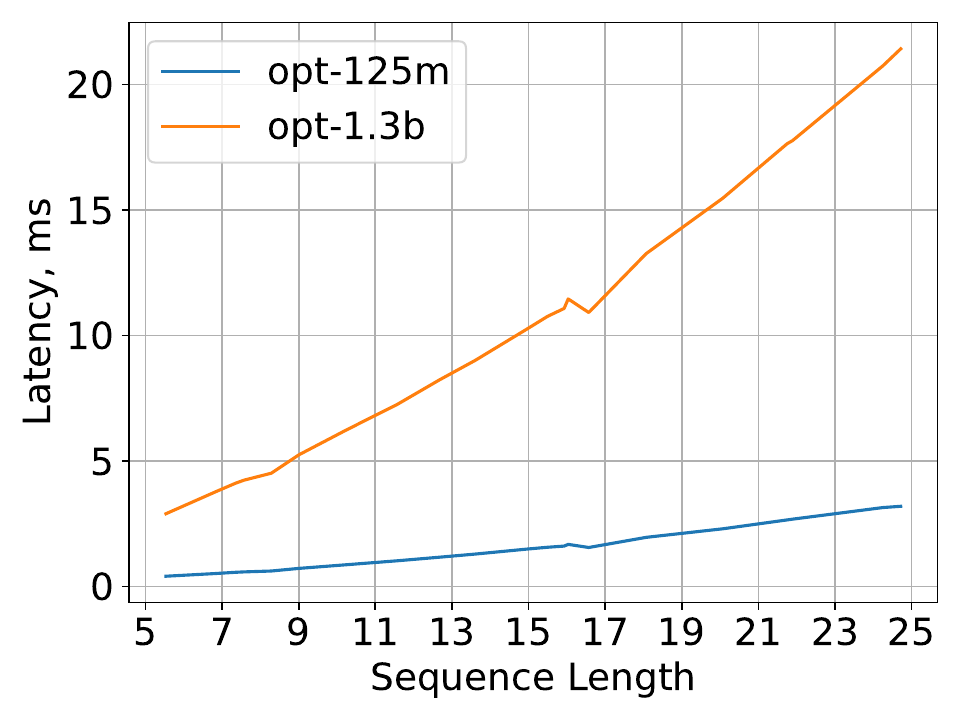}}
    }
    \caption{Inference Latency. Testing is conducted on a single A100-80G with the batch size set to the maximum possible value, the average number of tokens per item is 25.}
    \label{fig:latency}
    
\end{figure}

\begin{figure}[htbp]
    \centering{
    \subfigure{\includegraphics[width=0.8\textwidth]{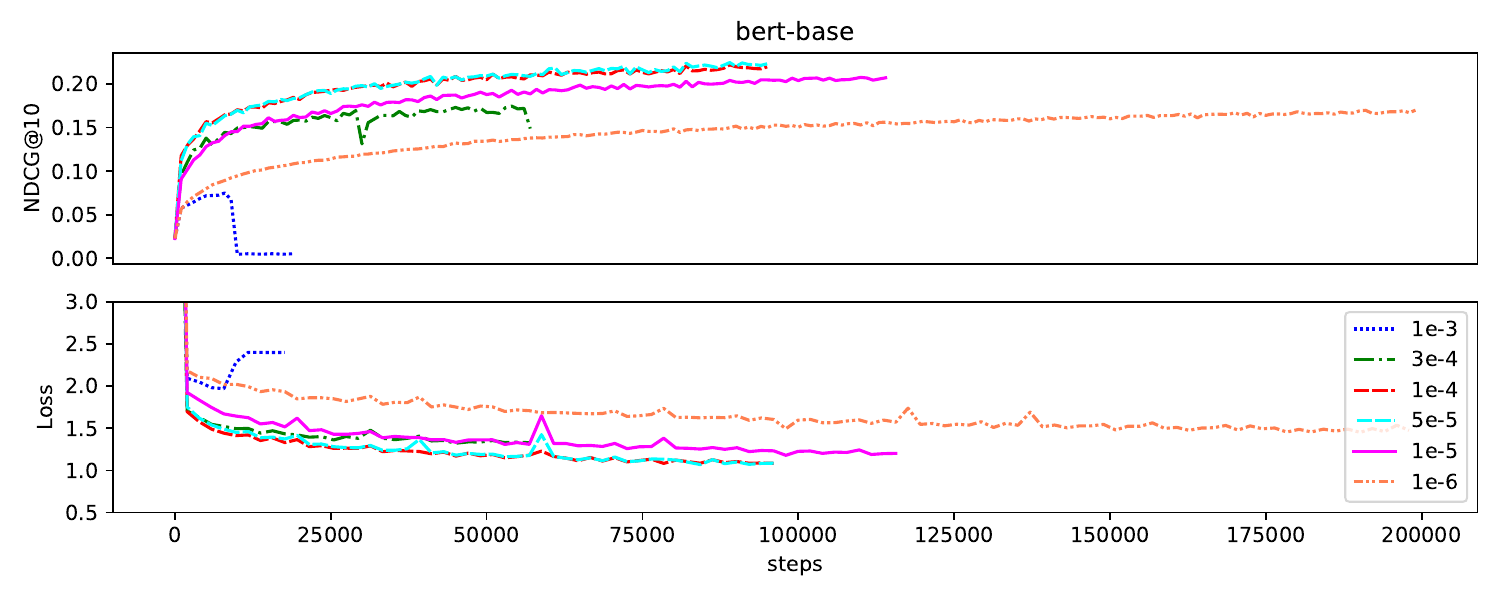}}
    \subfigure{\includegraphics[width=0.8\textwidth]{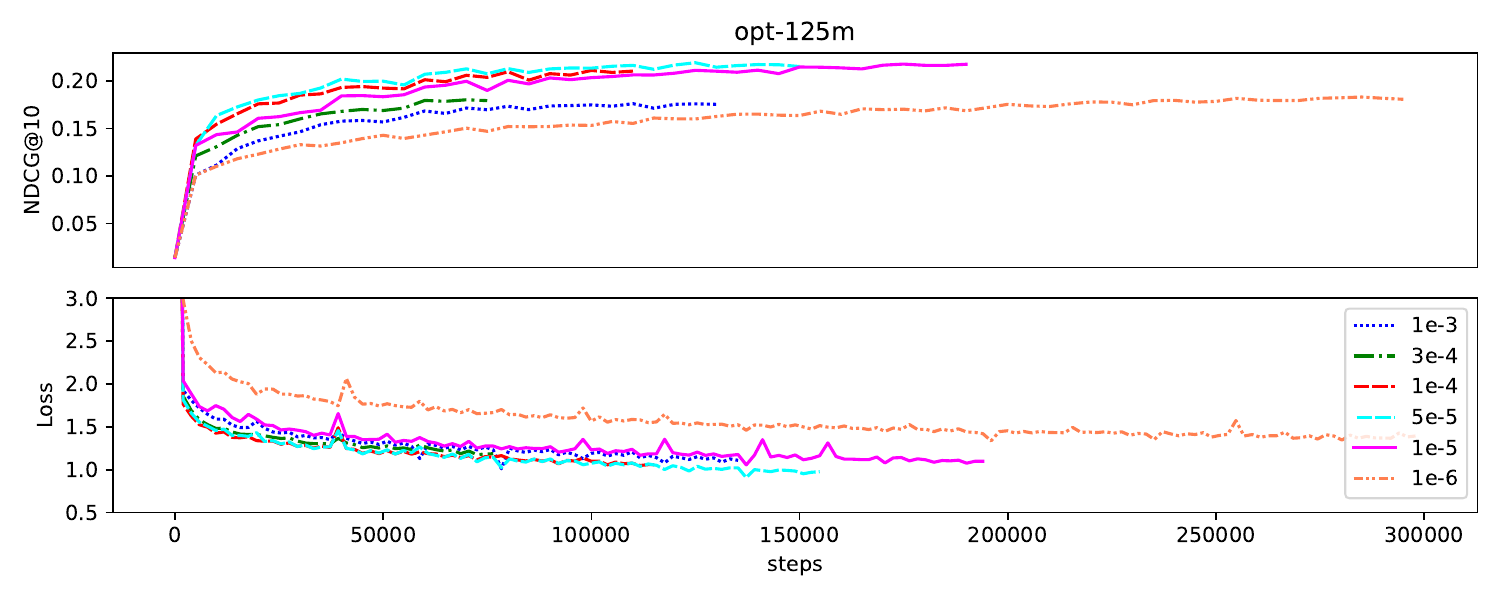}}
    }
    \caption{Performance of validation set and training loss trends across various learning rates during training. Each learning rate is represented by one color, with performance depicted as an overall upward trend in the curve, and training loss illustrated as an overall downward trend in the curve.}
    \label{fig:hyper}
\end{figure}

\subsubsection{Deployment Cost}
\paragraph{Training Cost}
Understanding the training cost is crucial for assessing the practicality and efficiency of one method, especially for resource-consuming LLM-based methods. We, hereby, present the training time of the SASRec and different size OPT models on the SPIAO dataset. For LLM which is larger than 1B, we evaluate the time both under FPFT and PEFT. Additionally, we set the batch size to $1$ and investigate the minimum memory resource consumption (video memory, VMEM) of FPFT and PEFT (LoRA) under different large model sizes, as shown in the left part of Table~\ref{tab:training_cost}. From the results, we have the following observations: First, given that the model size of LLM significantly surpasses that of the conventional method, the training cost will be an inevitable challenge and bottleneck of model optimization. Second, we can observe that it is infeasible to FPFT OPT-6.7B model with a single A100-80G GPU. Third, compared with FPFE, PEFT occupies less VMEM, while FPFT obtains faster convergence speed if there are sufficient computational resources. We believe this is because the pre-training corpus of the language model and the recommendation corpus are quite different, thus, it will require sufficient time to fine-tune a very small number of parameters with LoRA for convergence. 

\paragraph{Inference Cost}
Taking into account the high sensitivity to latency in recommendation systems, we compare the inference latency between language models and traditional sequential methods, as shown in the right part of Table ~\ref{tab:training_cost}. We calculate the latency of the model average inference for one example with the batch size set to 1 and the batch size set to the maximum possible value. The data is randomly sampled from the test set and the test is conducted on a single A100-80G. It can be observed that \baby significantly slows down inference compared to traditional methods.
Moreover, due to the proliferation of user sequence length ensuing from the integration of multi-domain data, we herein present the inference latency exhibited by language models of two sizes at different sequence lengths, as shown in Figure ~\ref{fig:latency}. Our findings are as follows: (1) Within the confines of a certain sequence length, the inference latency exhibits a linear relationship with the sequence length; (2) Increasing the sequence length on larger models results in a quicker rise in latency, with a slope significantly steeper than that of smaller models; 
Combining the results from Table ~\ref{tab:training_cost} Right and Figure ~\ref{fig:latency}, we can conclude that employing small models (around 100M) for sequence modeling is viable. With a sequence length of 25 and approximately 650 tokens, the encoding latency for a single sequence is around 3 milliseconds. However, when using large models (bigger than 1B) for inference, collaboration with acceleration strategies becomes necessary to avoid excessive latency.

\paragraph{Hyper-parameter Analysis.}
It is known that the choice of learning rate has a significant impact on performance when fine-tuning pre-trained language models. Therefore, we visualize the training process for two backbones under different learning rates. The performance patterns are illustrated in Figure ~\ref{fig:hyper}. From the results, we can observe that for the BERT-based model, a higher learning rate may cause the training process to oscillate or even collapse. For instance, with a learning rate of 1e-3, the loss surges after several thousand steps and the performance drops to zero. At a learning rate of 3e-4, there are two significant sudden declines in both loss and performance. When the learning rate is too small(1e-6 or 1e-5), the model struggles to converge, leading to sub-optimal performance. As for the OPT model, although there is no collapse or oscillation at higher learning rates (e.g., 1e-3 or 3e-4), both too high and too small learning rates significantly affect the performance.

{
\begin{table}[htbp]
\setlength\tabcolsep{3pt}
\centering
\caption{Feature ablation. The metric is Recall@10.}
\begin{tabular}{cc|rrrrrr}
\bottomrule
Size                            & Variants      & Scientific    & Pantry    & Instruments   & Arts      & Office  & SPIAO      \\  \hline
\multirow{3}{*}{OPT-125M}       & \baby         & 24.26         & 18.92     & 28.91         & 35.01     & 32.02   & 30.36      \\
                                & + id          & 25.17         & 21.11     & 30.33         & 36.38     & 32.83   & 31.50      \\
                                & + prompt      & 23.94         & 18.88     & 28.88         & 34.40     & 31.43   & 29.90       \\ \hline
\multirow{3}{*}{OPT-1.3B}       & \baby         & 25.06         & 19.15     & 29.16         & 35.48     & 32.15   & 30.69      \\
                                & + id          & \textbf{26.62}& \textbf{22.02} & \textbf{31.25} & \textbf{37.50} & \textbf{33.33}   & \textbf{32.33}  \\
                                & + prompt      & 24.29         & 18.71     & 28.92         & 35.22     & 31.87   & 30.30    \\

\toprule
\end{tabular}
\label{tab:feature_ablation}
\end{table}
}

\subsection{Feature ablation}
Although we primarily aim to investigate the application of language models and textual information for the multi-domain recommendation, the ID—an attribute that accurately denotes an item—emerges as an important feature. We explore the \baby's performance by integrating both ID and textual characteristics for item representation. In this case, we prepend the ID of the item to the beginning of the item title: $ID:id_i, t_v$. Furthermore, prompting is a crucial technical for adapting LLMs for downstream tasks. While we use LLMs for generating representations instead of dialogue, and considering the extensive fine-tuning on domain-specific data, a fixed and simple prompt may not be useful, so we use a fixed prompt to validate our assumption. In this case, we prepend a prompt sentence to the beginning of the user text sequence, and the prompt we use is \textit{"The user has purchased the following products in chronological order. Please select the next product they may interact with from the candidate products: "}. The results are shown in Table ~\ref{tab:feature_ablation}. From the results, we find that the ID feature can further improve the performance of \baby, whether on the small model (OPT-125M) or the large model (OPT-1.3B), this indicates that language models could capture the precise information represented by Id. The results of the prompt align with our speculation. As we use LLMs for generating representations and fine-tuning LLMs with domain-specific data, the simple prompt does not facilitate improvement. On the contrary, it may lead to a decrease in performance, as it can be regarded as meaningless noise.

\section{Conclusion}
\label{sec:conclusion}
In this paper, we propose a multi-domain sequential recommendation framework based on the large language model to alleviate the data sparsity and cold-start problems. By representing items and users with item titles and concatenation of interacted item titles, we utilize a pre-trained language model for generating corresponding representations. Furthermore, we mix user interactions across different domains to capture multi-domain knowledge. The experimental results on five real-world datasets demonstrate the effectiveness of \baby and the multi-domain information. 
Moreover, we study whether language model recommendations rely on comprehension or memory by analyzing cold-start performance, the Matthew effect, and multi-domain visualization. The experiments indicate that while there is a certain level of understanding for cold-start items, the recommendations rely more heavily on the memory of item-level collaborative filtering information.
Additionally, we investigate the impact of model size and tuning methods when applying large language models in recommendation. The experimental results show that as the size of the pre-trained language model increases, the fine-tune domain recommendation performance will increase slightly, while the zero-shot domain recommendation performance increase significantly. Moreover, the PEFT method fails to achieve promising performance when its model size is relatively small, and FPFT method is more effective given sufficient computational resources.

\begin{acks}
    This work was supported by National Natural Science Foundation of China (No.~62272349); Young Top-notch Talent Cultivation Program of Hubei Province; CCF-AFSG Research Fund; Ant Group through CAAI-Ant Research Fund. Lixin Zou and Chenliang Li are the corresponding authors.
\end{acks}

\bibliographystyle{ACM-Reference-Format}
\bibliography{bib}

\end{document}